\documentstyle[12pt,epsfig,amsbsy]{article}
\makeatletter

\newcommand{\PRD}[1]{{\it Phys.\ Rev.\ }{\bf D#1}}
\newcommand{\PLB}[1]{{\it Phys.\ Lett.\ }{\bf B#1}}
\newcommand{\PRC}[1]{{\it Phys.\ Rep.\ }{\bf C#1}}
\newcommand{\NPB}[1]{{\it Nucl.\ Phys.\ }{\bf B#1}}
\newcommand{\ZPC}[1]{{\it Z.\ Phys.\ }{\bf C#1}}
\newcommand{\CPC}[1]{{\it Comput.\ Phys.\ Commun.\ }{\bf #1}}
\newcommand{\hep}[1]{[{\rm hep-ph/#1}]}

\newcommand{\bep}{\boldsymbol{\mathrm{e}}^{\boldsymbol{+}}}
\newcommand{\bem}{\boldsymbol{\mathrm{e}}^{\boldsymbol{-}}}

\newcommand{\mrm}[1]{\mathrm{#1}}
\newcommand{\upp}{{\mrm{upp}}}
\newcommand{\low}{{\mrm{low}}}

\renewcommand{\d}{{\mrm{d}}}
\newcommand{\e}{{\mrm{e}}}

\newcommand{\p}{{\mrm{p}}}

\newcommand{\pbar}{\overline{{\mrm{p}}}}

\newcommand{\me}{m_{\e}}

\newcommand{\lessim}{\raisebox{-0.8mm}%
{\hspace{1mm}$\stackrel{<}{\sim}$\hspace{1mm}}}

{\end{list}}
\newcounter{enumct}

\newlength{\abstwidth}
\def\citenum#1{{\def\@cite##1##2{##1}\cite{#1}}}
\def\citea#1{\@cite{#1}{}}
 
\newcount\@tempcntc
\def\@citex[#1]#2{\if@filesw\immediate\write\@auxout{\string
\citation{#2}}\fi
  \@tempcnta\z@\@tempcntb\m@ne\def\@citea{}\@cite{\@for\@citeb:=#2\do
    {\@ifundefined
       {b@\@citeb}{\@citeo\@tempcntb\m@ne\@citea\def\@citea{,}{\bf ?}
\@warning
       {Citation `\@citeb' on page \thepage \space undefined}}%
    {\setbox\z@\hbox{\global\@tempcntc0\csname b@\@citeb\endcsname
\relax}%
     \ifnum\@tempcntc=\z@ \@citeo\@tempcntb\m@ne
       \@citea\def\@citea{,}\hbox{\csname b@\@citeb\endcsname}%
     \else
      \advance\@tempcntb\@ne
      \ifnum\@tempcntb=\@tempcntc
      \else\advance\@tempcntb\m@ne\@citeo
      \@tempcnta\@tempcntc\@tempcntb\@tempcntc\fi\fi}}\@citeo}{#1}}
\def\@citeo{\ifnum\@tempcnta>\@tempcntb\else\@citea\def\@citea{,}%
  \ifnum\@tempcnta=\@tempcntb\the\@tempcnta\else
   {\advance\@tempcnta\@ne\ifnum\@tempcnta=\@tempcntb \else \def
\@citea{--}\fi
    \advance\@tempcnta\m@ne\the\@tempcnta\@citea\the\@tempcntb}\fi\fi}

\begin{document}
%set sloppy attitude to line breaks
\sloppy

\pagestyle{empty}

\begin{flushright}
CERN--TH/96--313\\
hep-ph/9611249
\end{flushright}
 
\vspace{\fill}
 
\begin{center}
{\Large{\bf  
GALUGA, a Monte Carlo program for\\[1ex]
two-photon processes in}
$\bep\bem$ {\bf collisions}}\\[10mm]
{\Large Gerhard A. Schuler$^a$} \\[3mm]
{\it Theory Division, CERN,} \\[1mm]
{\it CH-1211 Geneva 23, Switzerland}\\[1mm]
{ E-mail: Gerhard.Schuler@cern.ch}
\end{center}
 
\vspace{\fill}
 
\begin{center}
{\bf Abstract}\\[2ex]
\begin{minipage}{\abstwidth}
A Monte Carlo program is presented for the computation of the most 
general cross section for two-photon production in $\e^+\e^-$ collisions 
at fixed two-photon invariant mass $W$. Functions implemented for 
the five $\gamma^\star\gamma^\star$ structure functions include 
three models of the total hadronic cross section and the lepton-pair 
production cross section. Prospects of a structure-function determination 
through a study of the azimuthal dependence between the two scattering 
planes are outlined. All dependences on the electron mass and the photon 
virtualities $Q_i^2$ are fully kept. Special emphasis is put on a 
numerically stable evaluation of all variables over the full $Q_i^2$ 
range from $Q_{i\min}^2 \sim \me^2 (W/\sqrt{s})^4 \ll \me^2$ up to 
$Q_{i\max}^2 \sim s$. A comparison is made with an existing 
Monte Carlo program for lepton-pair production and an 
equivalent-photon approximation for hadronic cross sections.
\end{minipage}
\end{center}

\vspace{\fill}
\noindent
\rule{60mm}{0.4mm}

\vspace{1mm} \noindent
${}^a$ Heisenberg Fellow.

\vspace{10mm}\noindent
CERN--TH/96--313 \\
October 1996

\clearpage
\pagestyle{plain}
\setcounter{page}{1} 

\begin{center}
  \begin{large}
     Program Summary
  \end{large}
\end{center}

\renewcommand{\arraystretch}{1.1}

\noindent
\begin{tabular}{p{0.54\textwidth}p{0.40\textwidth}}
{\it Title of program:} & GALUGA
\\
{\it Program obtainable from:} & G.A.\ Schuler,  CERN--TH, 
\\ & CH-1211 Geneva 23, Switzerland; 
\\ &Gerhard.Schuler@cern.ch
\\
{\it Licensing provisions:} & none
\\
{\it Computer for which the program is designed and others on which 
it is operable:} & all computers
\\
{\it Operating system under which the program has been tested:} & UNIX
\\
{\it Programming language used:} & FORTRAN 77
\\
{\it Number of lines:} & $1445$
\\
{\it Keywords:} & Monte Carlo, two-photon, $\e^+\e^-$, azimuthal dependence
\\
{\it Subprograms used:} & VEGAS \cite{VEGAS} (included, 229 lines)
\\                & RANLUX \cite{RANLUX} (included, 305 lines)
\\                & HBOOK \cite{HBOOK} and DATIME \cite{DATIME} for
\\                & the test program (367 lines)
\\
\end{tabular}
{\it Nature of physical problem:}\\
Hadronic two-photon reactions in a new energy domain are becoming 
accessible with LEP2. Unlike purely electroweak processes, hadronic 
processes contain dominant non-perturbative components 
parametrized by suitable structure functions, which are functions 
of the two-photon invariant mass $W$ and the photon virtualities $Q_1$
and $Q_2$. It is hence advantageous to have a Monte Carlo program that can 
generate events at fixed, user-defined values of $W$ and, optionally,  
at fixed values of $Q_i$. Moreover, at least one program with 
an exact treatment of both the kinematics and the dynamics over the whole 
range $m^2 \gg m^2 (W/\sqrt{s})^4  \lessim Q_i^2 \lessim s$ 
($m$ is the electron mass and $\sqrt{s}$ the $\e^+\e^-$ c.m.\ energy) 
is needed, (i) to check the various approximations used in other 
programs, and (ii) to be able to explore additional information on the 
hadronic physics, e.g.\ coded in azimuthal dependences. 

\noindent
{\it Method of solution:}\\
The differential cross section for $\e^+\e^- \rightarrow \e^+\e^- X$ 
at fixed two-photon invariant mass $W$ is rewritten in terms of four 
invariants with the photon virtualities $Q_i$ as the two outermost 
integration variables in order to simultaneously cope with antitag and 
tagged electron modes. Due care is taken of numerically stable expressions 
while keeping all electron-mass and $Q_i$ dependences. Special care is 
devoted to the azimuthal dependences of the cross section. Cuts on the 
scattered electrons are to a large extent incorporated analytically and
suitable mappings introduced to deal with the peaking structure of the 
differential cross section. The event generation yields either weighted 
events or unweighted ones (i.e.\ equally weighted events with weight $1$), 
the latter based on the hit-or-miss technique.
%by comparing the actual function value with a random number times the 
%maximum function value, which is also determined during the VEGAS call. 
Optionally, VEGAS can be invoked to (i) obtain an accurate estimate of the 
integrated cross section 
%within the cuts on the scattered electrons defined by the user 
and (ii) improve the event generation efficiency through additional 
variable mappings provided by the grid information of VEGAS. 
The program is set up so that additional hadronic (or leptonic) reactions 
can easily be added.

\noindent
{\it Typical running time:}\\
The integration time depends on the required cross-section accuracy and the 
applied cuts. For instance, $13$ seconds on an IBM RS/6000 yields an 
accuracy of the VEGAS integration of about $0.1$\% for the antitag mode 
or of about $0.2\%$ for a typical single-tag mode;  
within the same time the error of the  simple 
Monte Carlo integration is about $0.5$\% for either mode. 
Event generation with or without VEGAS improvement and for either tag mode  
takes about $4\times 10^{-4}$ ($2\times 10^{-3}$)
seconds per event for weighted (unweighted) events.

\renewcommand{\arraystretch}{1.5}
\clearpage
\section{Introduction}
Two-photon physics is facing a revival with the advent of LEP2. 
Measurements of two-photon processes in a new domain of $\gamma\gamma$ 
c.m.\ energies $W$ are ahead of us \cite{LEP2}. 
Any two-photon process is, in general, described \cite{Budnev} by five 
non-trivial structure functions (two more for polarized initial electrons). 
Purely QED (or electroweak) processes are fully calculable within
perturbation theory. Several sophisticated Monte Carlo event generators  
exist \cite{vermaseren,diag36,Diberder}
to simulate $4$-fermion production in $\e^+\e^-$ collisions. 
Indeed, the differential cross section is not explicitly decomposed 
as an expansion in the five $\gamma^\star\gamma^\star$ structure functions. 
Rather, the full matrix element for the reaction $\e^+\e^- \rightarrow 
\e^+\e^- \ell^+\ell^-$ is calculated as a whole, partly even including 
QED radiative corrections. 
Such a procedure is, however, not possible for hadronic two-photon reactions
since the hadronic behaviour of the photon is of non-perturbative 
origin. The decomposition into the above-mentioned five structure functions 
(and their specification, of course) is hence mandatory for a full 
description of hadronic reactions. 

Monte Carlo event generators for hadronic two-photon processes can
be divided into two classes. Programs of
the first kind \cite{herwig,pythia,phojet,minijet,ggps1,gghv01}
put the emphasis on the QCD part but are (so far) restricted to the 
scattering of two {\em real} photons. 
The two-photon sub-processes are then embedded in an approximate 
way in the overall reaction of $\e^+\e^-$ collisions. 
A recent discussion of the so-called equivalent-photon 
approximation can be found in \cite{GAS}. 

The other type of programs \cite{twogam,twogen} treat the 
kinematics of the vertex $\e^+\e^- \rightarrow \e^+\e^-\gamma\gamma$
more exactly, but they contain only simple models of the hadronic physics. 
Moreover, the event generation is done in the variables that are tailored 
for $\e\e\rightarrow \e\e\gamma\gamma$, namely the energies and angles
(or virtualities) of the photons and the azimuthal angle $\phi$ between
the two lepton-scattering planes in the {\em laboratory} system. 
Hence, both the hadronic energy $W$ and the azimuthal angle $\tilde{\phi}$ 
in the {\em photon} c.m.s.\ (which enters the decomposition of the
$\e^+\e^-$ cross section into the five hadronic structure functions) 
are highly non-trivial functions of these variables\footnote{The only 
program that contains the $\tilde\phi$-dependences 
is TWOGAM \cite{twogam}. However, the expressions taken from \cite{Budnev} 
are numerically very unstable at small $Q_i$; see the discussion following
(\ref{rhoidef}). Moreover, $\tilde{\phi}$ itself is not calculated.}.

In the study of hadronic physics one prefers to study events 
at fixed values of $W$. Not only is $W$ the crucial variable that 
determines the nature of the hadronic physics, but through studies of 
events at fixed $W$ can $\gamma\gamma$ collisions be compared with
$\gamma\p$ and $\p\pbar$ ones \cite{SaS}. Next to $W$, the virtualities 
$Q_1$ and $Q_2$ of the two photons determine the hadronic physics. 
At fixed values of $W$ and one of the $Q$'s, $Q_1$ say, one obtains 
the cross section of deep-inelastic electron--photon scattering. 
Varying $Q_2$ one can investigate the so-called target-mass effects,
i.e.\ the influence of non-zero values of $Q_2$ on the extraction of 
the photon structure function $F_2$. Hence it is desirable to have 
an event generator that keeps $W$ fixed and in which 
$Q_1$ and $Q_2$ are the outermost integration variables so that these
can be held constant. 

The remaining two non-trivial integration variables, which complete 
the phase space of $\e^+\e^- \rightarrow \e^+\e^- X$, should be chosen 
such that three conditions are fulfilled. First, cuts on the scattered 
electrons are usually imposed in experimental analyses. Hence, 
the efficiency and accuracy of the program is improved if 
these can be treated explicitly rather than incorporated by a 
simple rejection of those events that fall outside the allowed region. 
Second, the peaking structure of the differential cross section should
be reproduced as well as possible in order to reduce the estimated 
Monte Carlo error and to improve the efficiency of the event generation.
And third, it should be possible to achieve a numerically stable evaluation 
of all variables needed for a complete event description. These three
conditions are met to a large extent by the choice of subsystem 
squared invariant masses $s_1$ and $s_2$ as integration variables 
besides $Q_1^2$ and $Q_2^2$. In the laboratory frame, $s_i$ are related 
to the photon energies $\omega_i$ by $s_{1/2} - m^2 = 2 \omega_{2/1}
\sqrt{s}$, where $m$ denotes the electron mass.

In the interest of those readers not interested in calculational details, 
the paper starts with a presentation of a few results in 
section~\ref{sec:results}. The differential cross section for the reaction 
$\e^+\e^- \rightarrow \e^+\e^- X$ is rewritten in terms of the four 
invariants $Q_i^2$ and $s_i$ ($i=1,2$) in section~\ref{sec:notation} where 
also models for the cross section $\sigma(\gamma^\star\gamma^\star 
\rightarrow X)$ are described. 
The integration boundaries with $Q_1^2$ and $Q_2^2$ as the two outermost 
integration variables are specified in section~\ref{sec:phase}. 
The derivation of the integration limits is standard \cite{Byckling} but 
tedious. Here the emphasis is put on numerically stable 
expressions\footnote{A similar phase-space decomposition 
with $s_1$ replaced by $\Delta = [(s-2m^2) (W^2+Q_1^2+Q_2^2)
-(s_1+Q_2^2-m^2)(s_2+Q_1^2-m^2)]/4$ is presented in \cite{vermaseren}.}. 
To our knowledge, numerical stable forms of $\phi$ and $\tilde{\phi}$ 
are presented here for the first time. All dependences on the electron 
mass and the virtualities of the two photons are kept. The formulas
are stable over the whole range from $Q_{i\min}^2 \sim m^2 (W/\sqrt{s})^4 
\ll m^2$ up to $Q_{i\max}^2 \sim s$, i.e.\ the program covers smoothly 
the antitag and tag regions. An equivalent-photon approximation is
also implemented (section~\ref{sec:EPA}). 
The complete representation of the four-momenta of the produced particles 
in terms of the integration variables is given in section~\ref{sec:momenta}. 
Section~\ref{sec:cuts} describes the incorporation of cuts on the scattered  
electrons. Details of the Monte Carlo program GALUGA 
are given in section~\ref{sec:program}.

\section{A few results}
\label{sec:results}
\begin{figure}
 \begin{center}
\begin{tabular}[t]{cc}
\epsfig{file=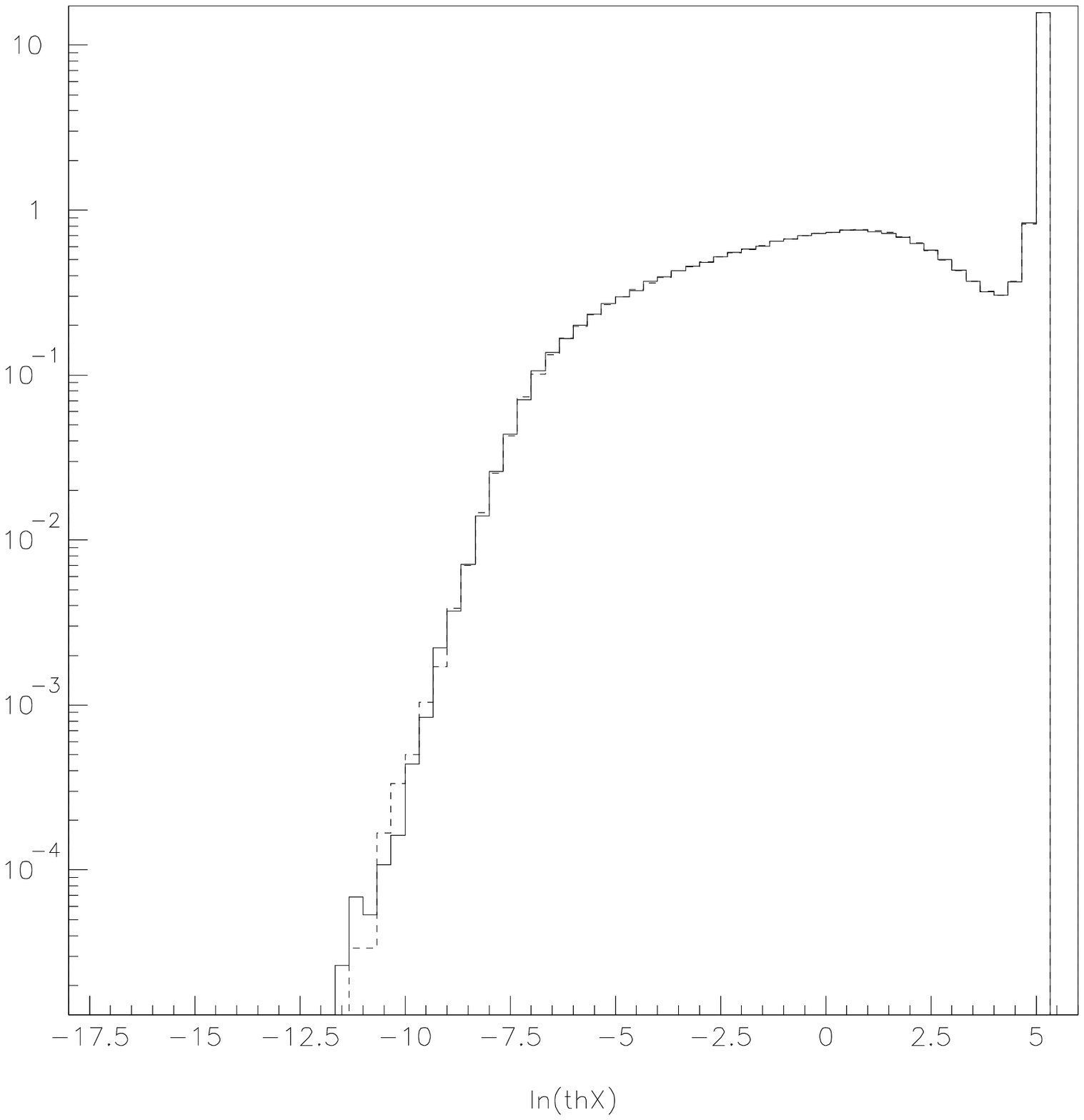,height=0.40\textheight}\\
\epsfig{file=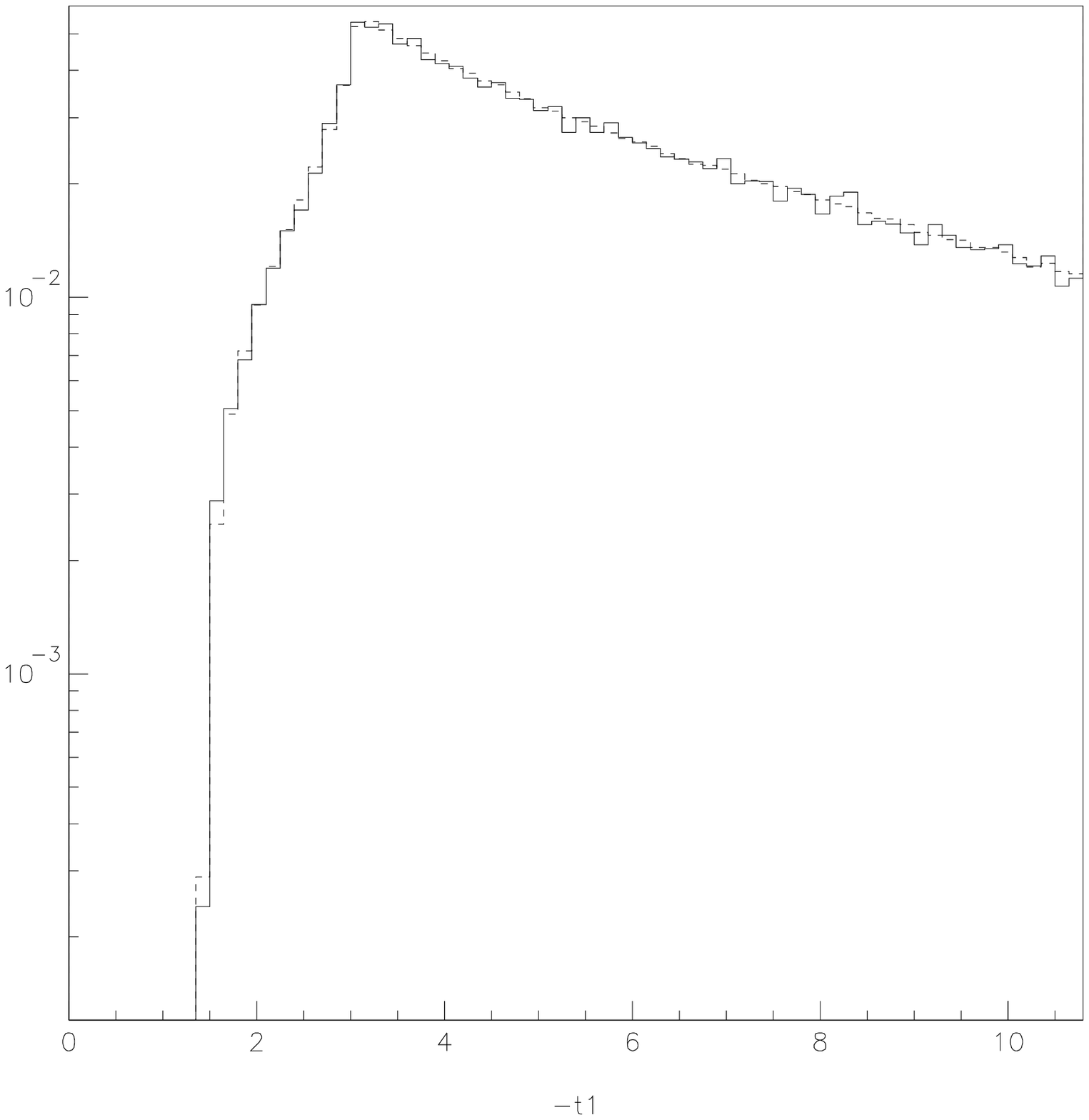,height=0.40\textheight}
\end{tabular}
\caption[]{Comparison of muon-pair production in   
GALUGA (dashed histograms) and DIAG36 (solid histograms) at
$\sqrt{s}=130\,$GeV and $W=10\,$GeV. Top: distribution in the
logarithm of the polar angle of the $\mu^+\mu^-$ system; 
no cuts are applied on the scattered electrons.
Bottom:  distribution in $t_1$ under the cuts: 
$1.55 < \theta_1 < 3.67^\circ$ and $30\,$GeV$ < E_1$.
\label{fig:blnthx}}
\end{center}
\end{figure}

In order to check GALUGA, we include the production of lepton pairs, 
for which several well-established Monte Carlo generators 
\cite{vermaseren,diag36,Diberder} exist. The five structure functions for 
$\gamma^\star\gamma^\star \rightarrow \ell^+ \ell^-$ 
as quoted in \cite{Budnev} have been implemented. 
For the comparison we have modified the two-photon part of the 
four-fermion program DIAG36 \cite{diag36} 
(i.e.\ DIAG36 restricted to the multiperipheral diagrams) 
in such a way that it can produce events at {\em fixed} values of $W$. 
The agreement is excellent. Two examples are shown 
in Fig.~\ref{fig:blnthx}, the first corresponding to a no-tag setup
and the second to a single tagging mode. 

\begin{figure}
 \begin{center}
\begin{tabular}[t]{cc}
   \epsfig{file=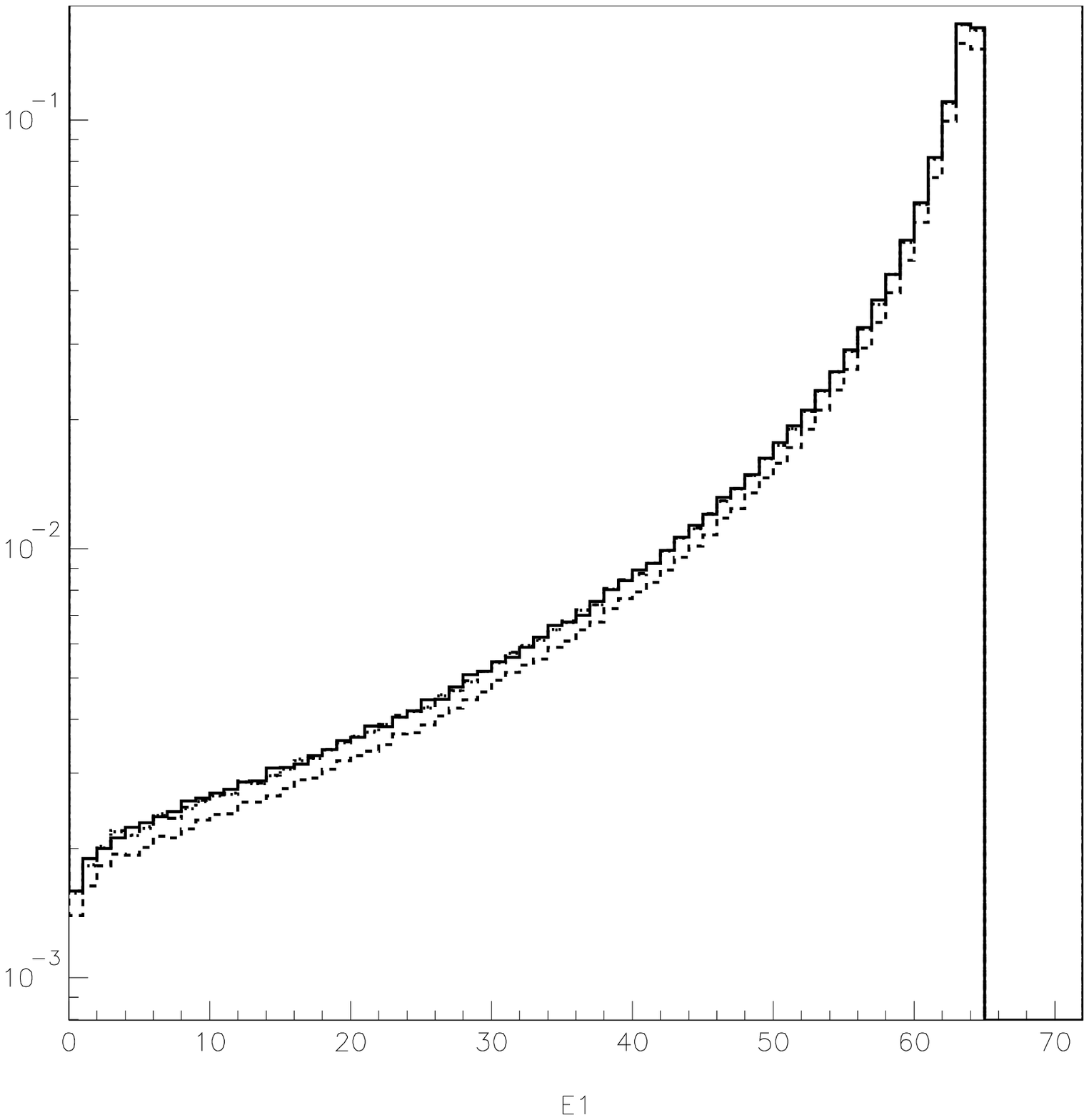,width=0.45\textwidth} &
   \epsfig{file=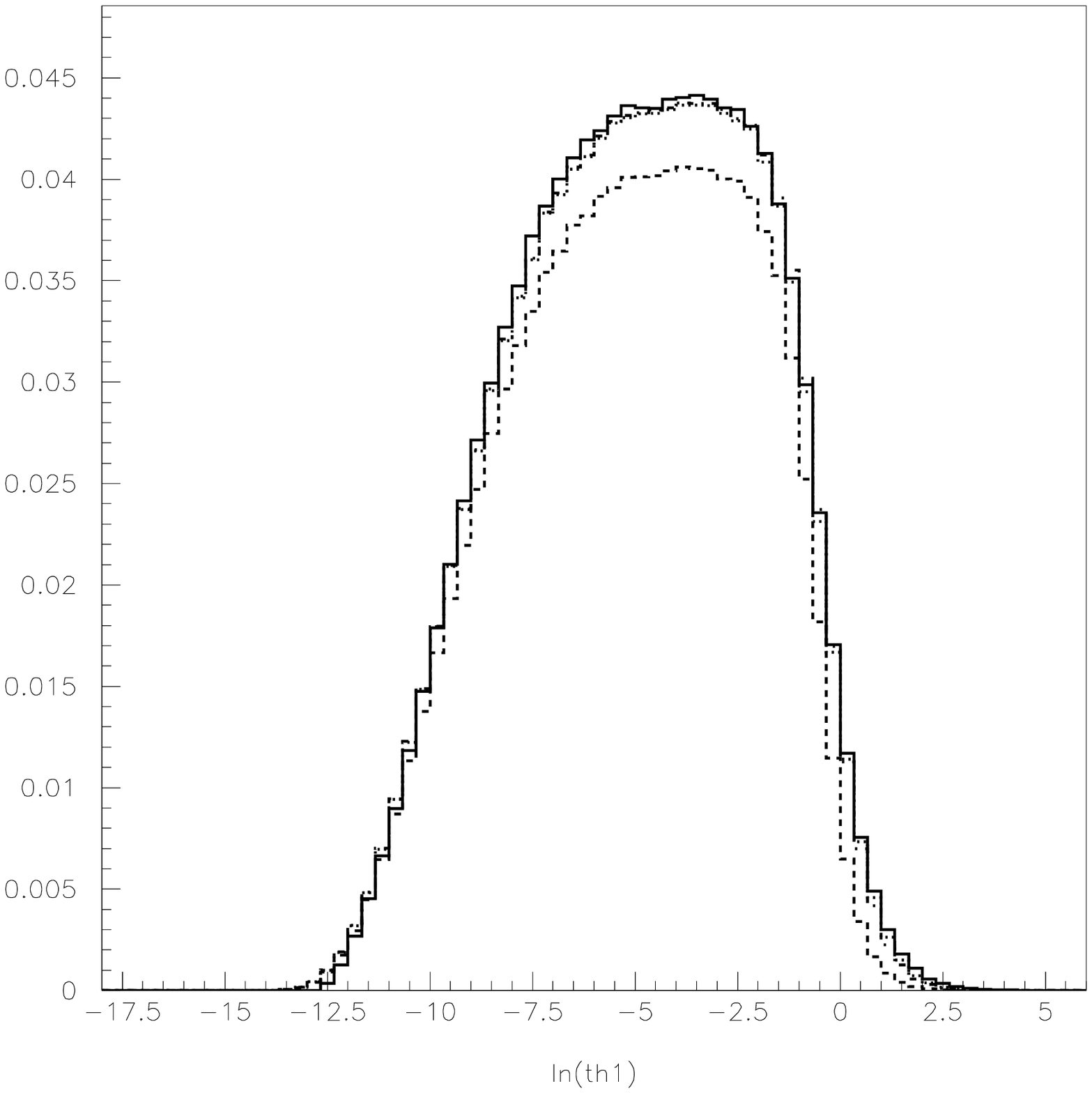,width=0.45\textwidth}
 \\
   \epsfig{file=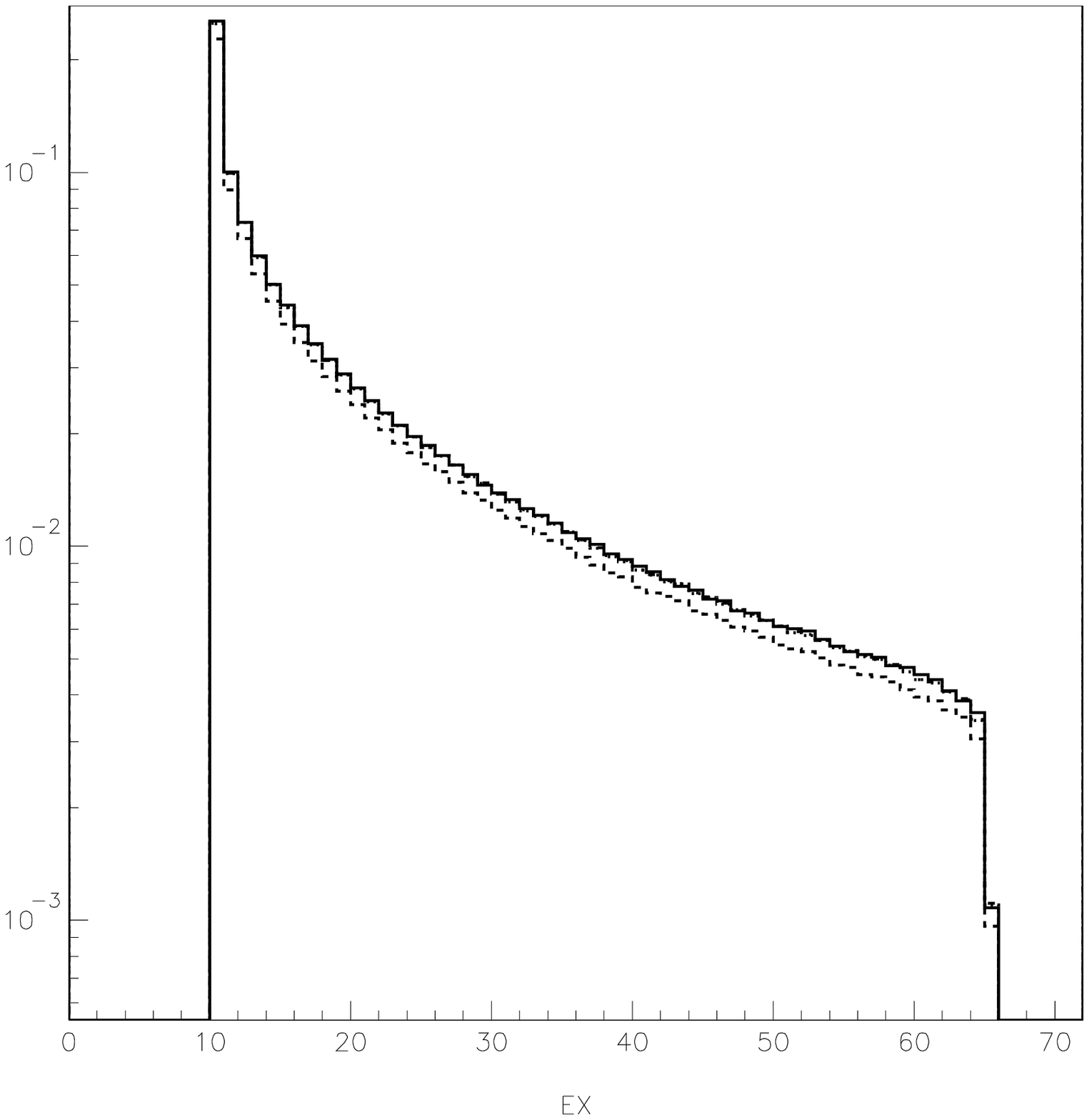,width=0.45\textwidth} &
   \epsfig{file=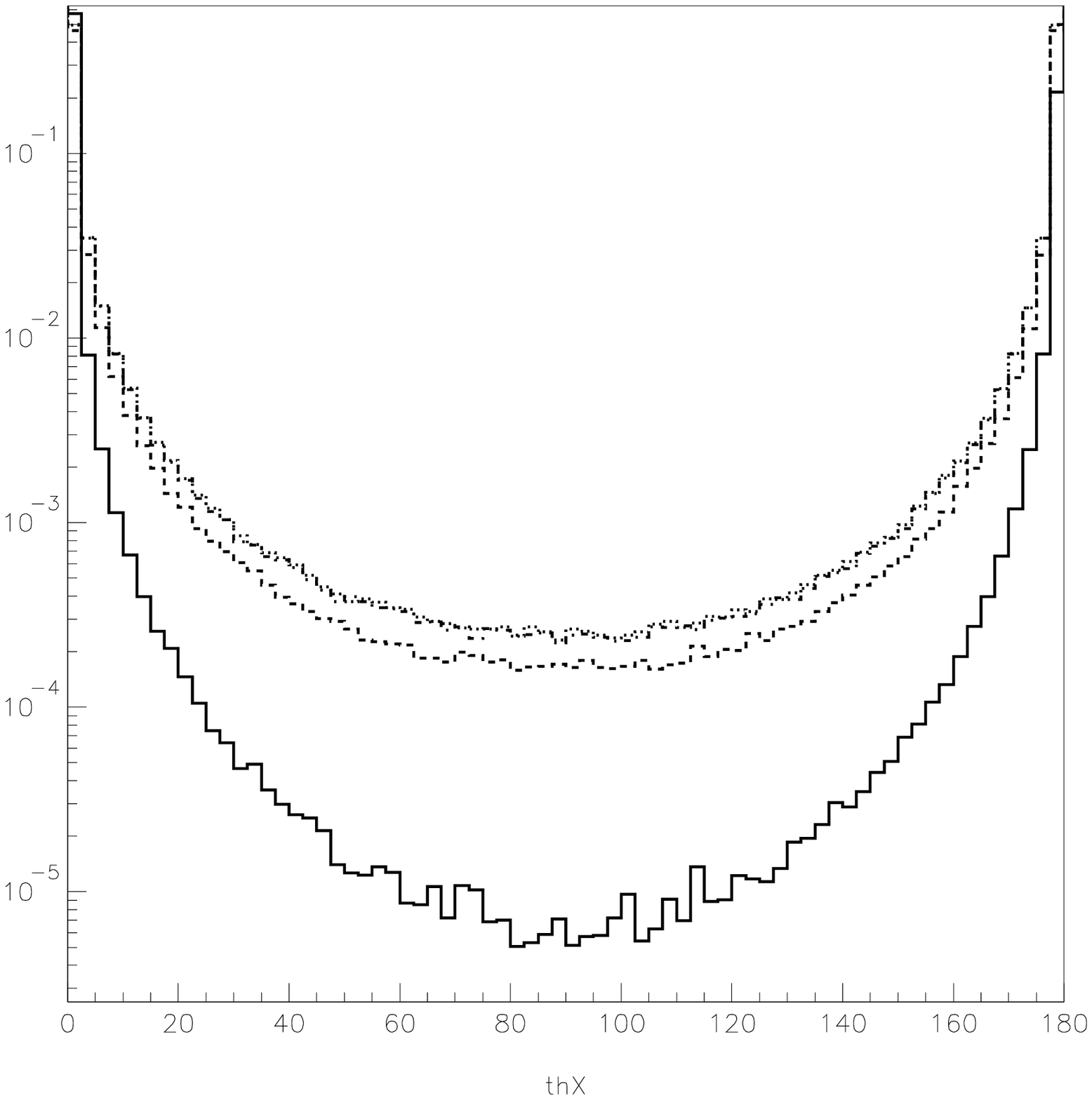,width=0.45\textwidth}
  \end{tabular}
\caption[]{Distributions in $E_1$, $\ln\theta_1$, $E_X$, and
$\theta_X$ for the integrated total hadronic cross section
at $\sqrt{s}=130\,$GeV and $W=10\,$GeV. No cuts on the scattered 
electrons are applied. Histogram line-styles correspond to GVMD model 
in the EPA (solid), $\rho$-pole model (dashed), GVMD model (dash-dotted), 
VMDc model (dotted).
\label{fig:nocuts}}
 \end{center}
\end{figure}

\begin{figure}
 \begin{center}
\begin{tabular}[t]{cc}
   \epsfig{file=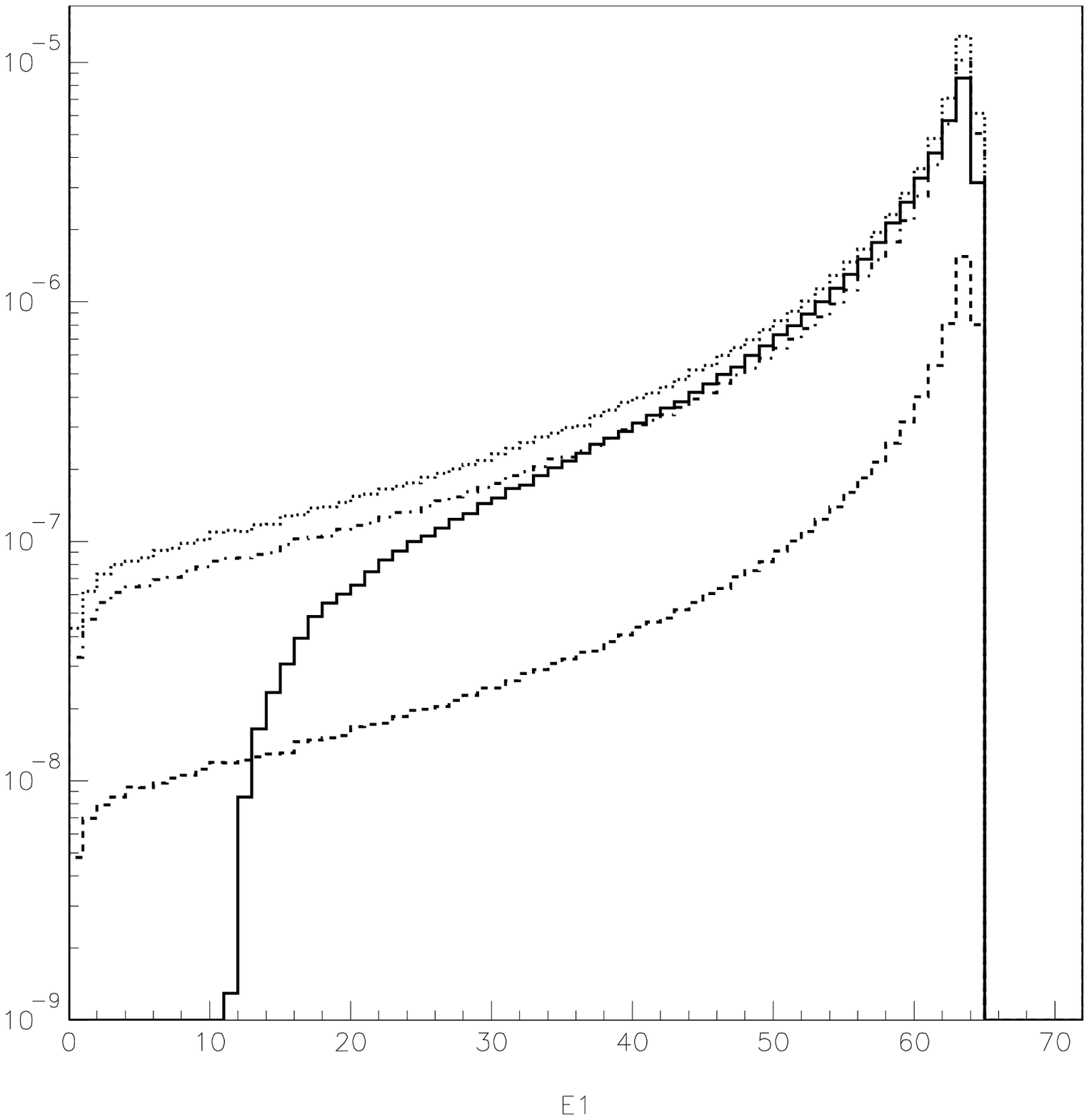,width=0.45\textwidth} &
   \epsfig{file=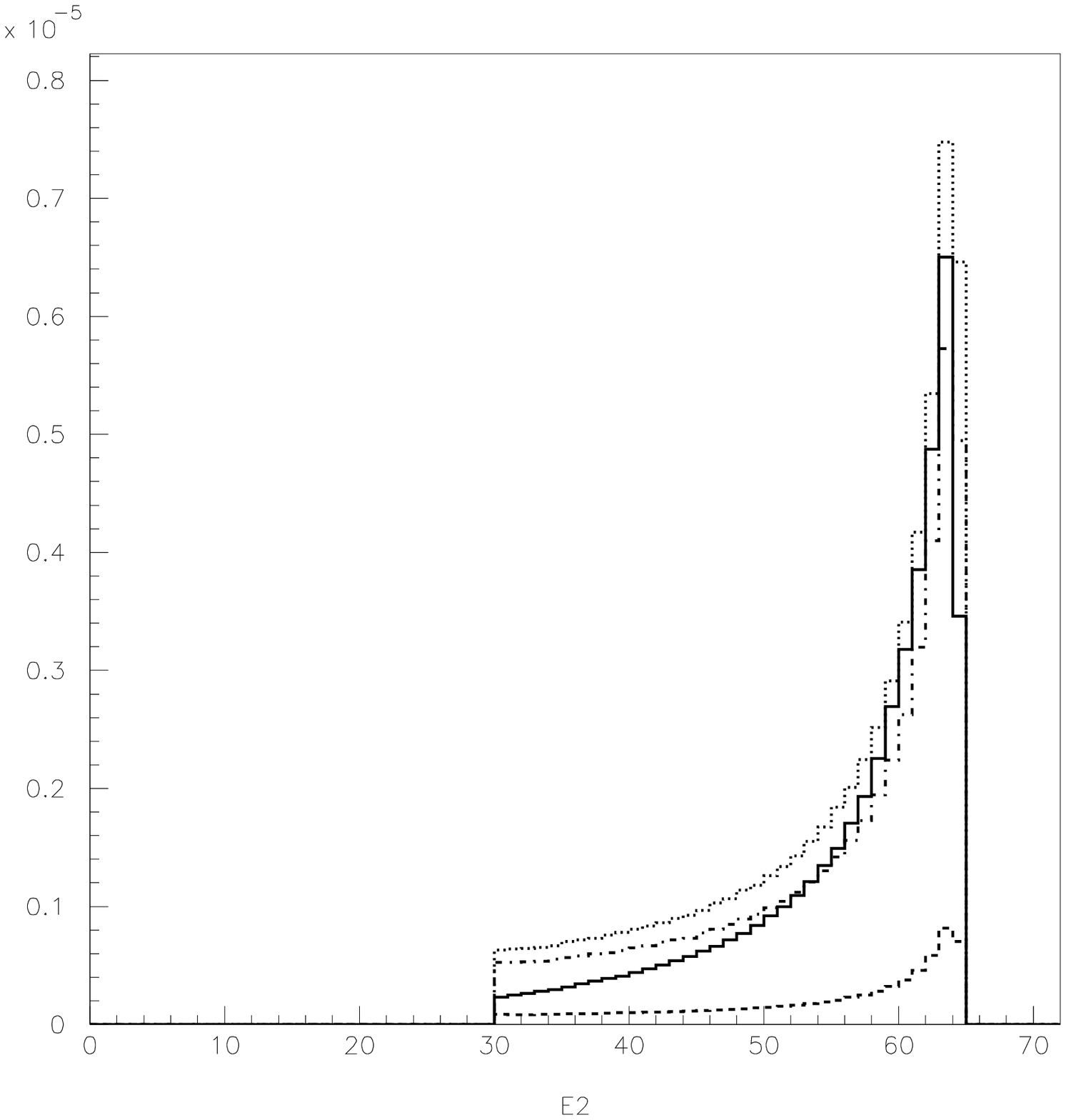,width=0.45\textwidth}
 \\
   \epsfig{file=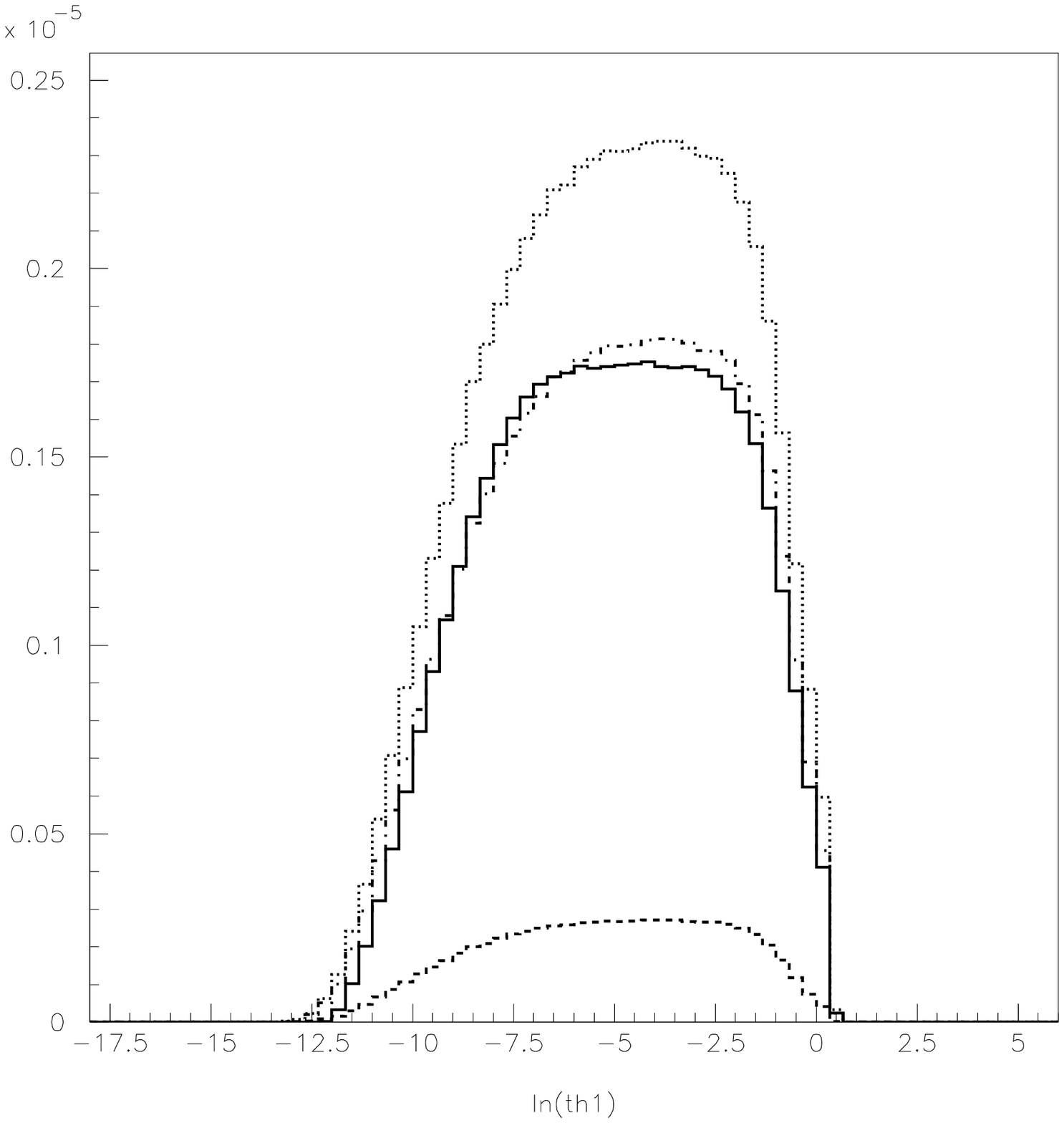,width=0.45\textwidth} &
   \epsfig{file=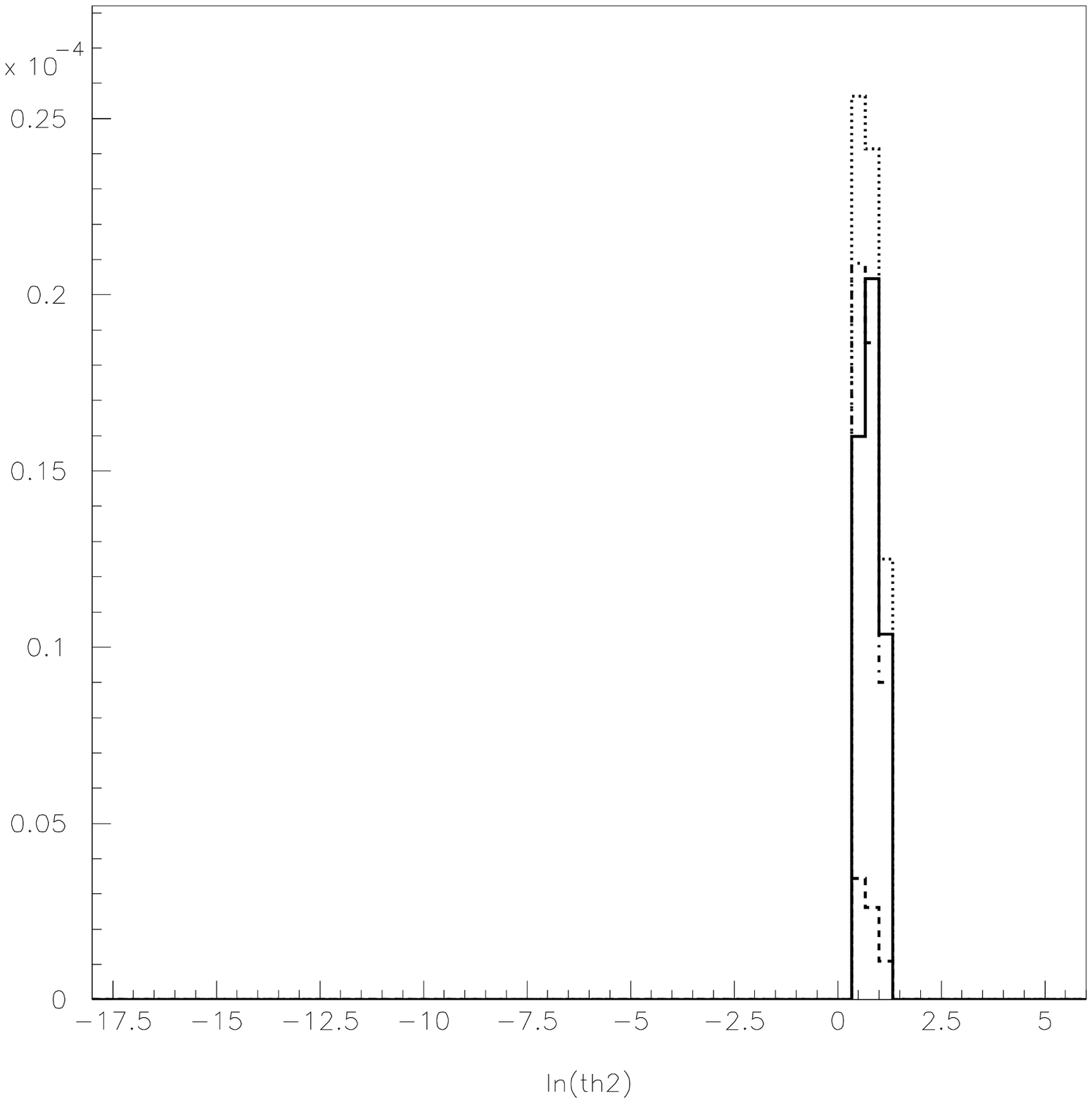,width=0.45\textwidth}
  \end{tabular}
\caption[]{Distributions in $E_1$, $E_2$, $\ln\theta_1$, and
$\ln\theta_2$ for the integrated total hadronic cross section
at $\sqrt{s}=130\,$GeV and $W=10\,$GeV. 
The cuts $\theta_1 < 1.43^\circ$, $1.55 < \theta_2 < 3.67^\circ$, 
and $30\,$GeV$< E_2$ have been applied.
Histogram line-styles correspond to GVMD model 
in the EPA (solid), $\rho$-pole model (dashed), GVMD model (dash-dotted), 
VMDc model (dotted).
\label{fig:cutone}}
 \end{center}
\end{figure}

\begin{figure}
 \begin{center}
\begin{tabular}[t]{cc}
   \epsfig{file=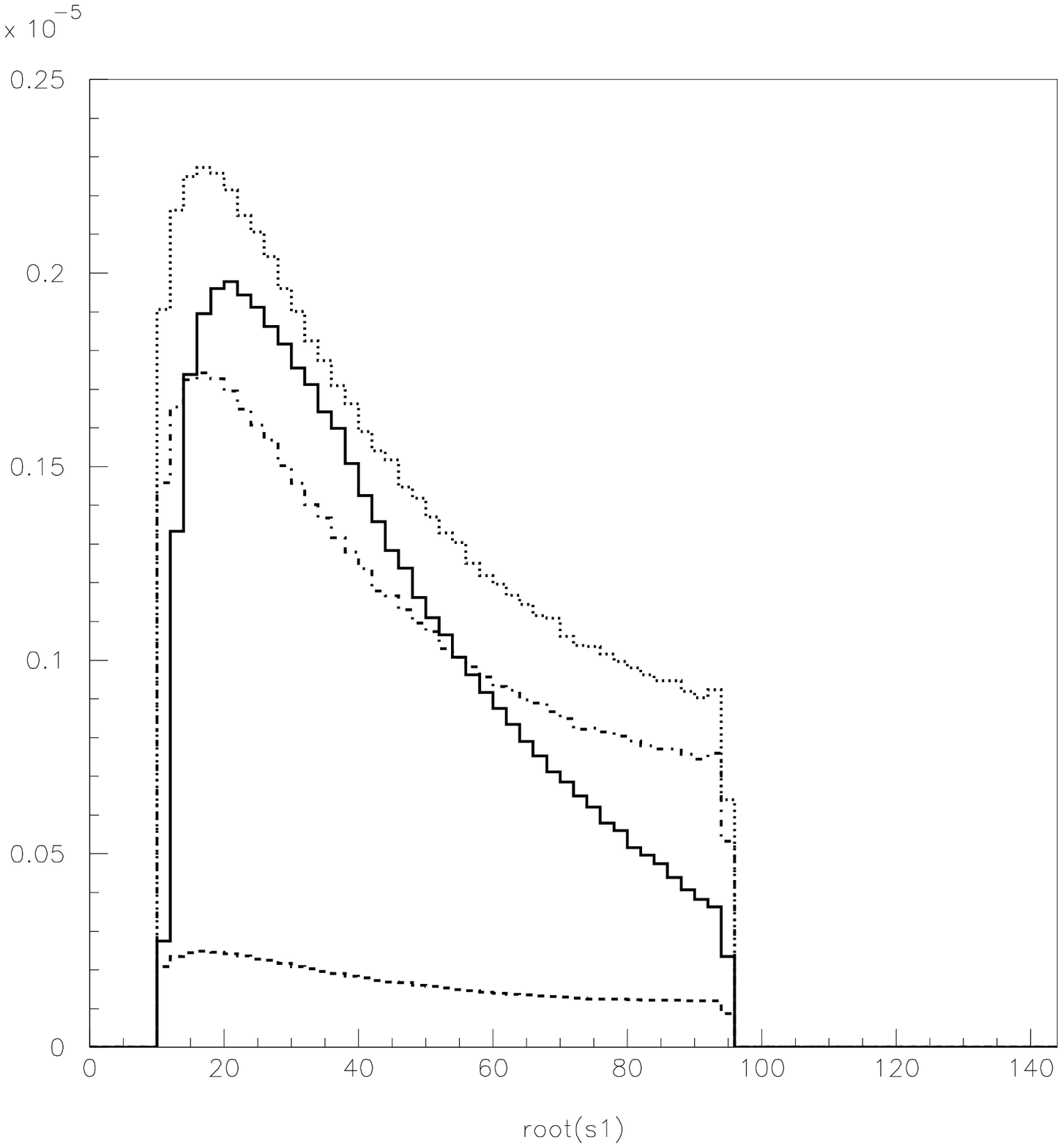,width=0.45\textwidth} &
   \epsfig{file=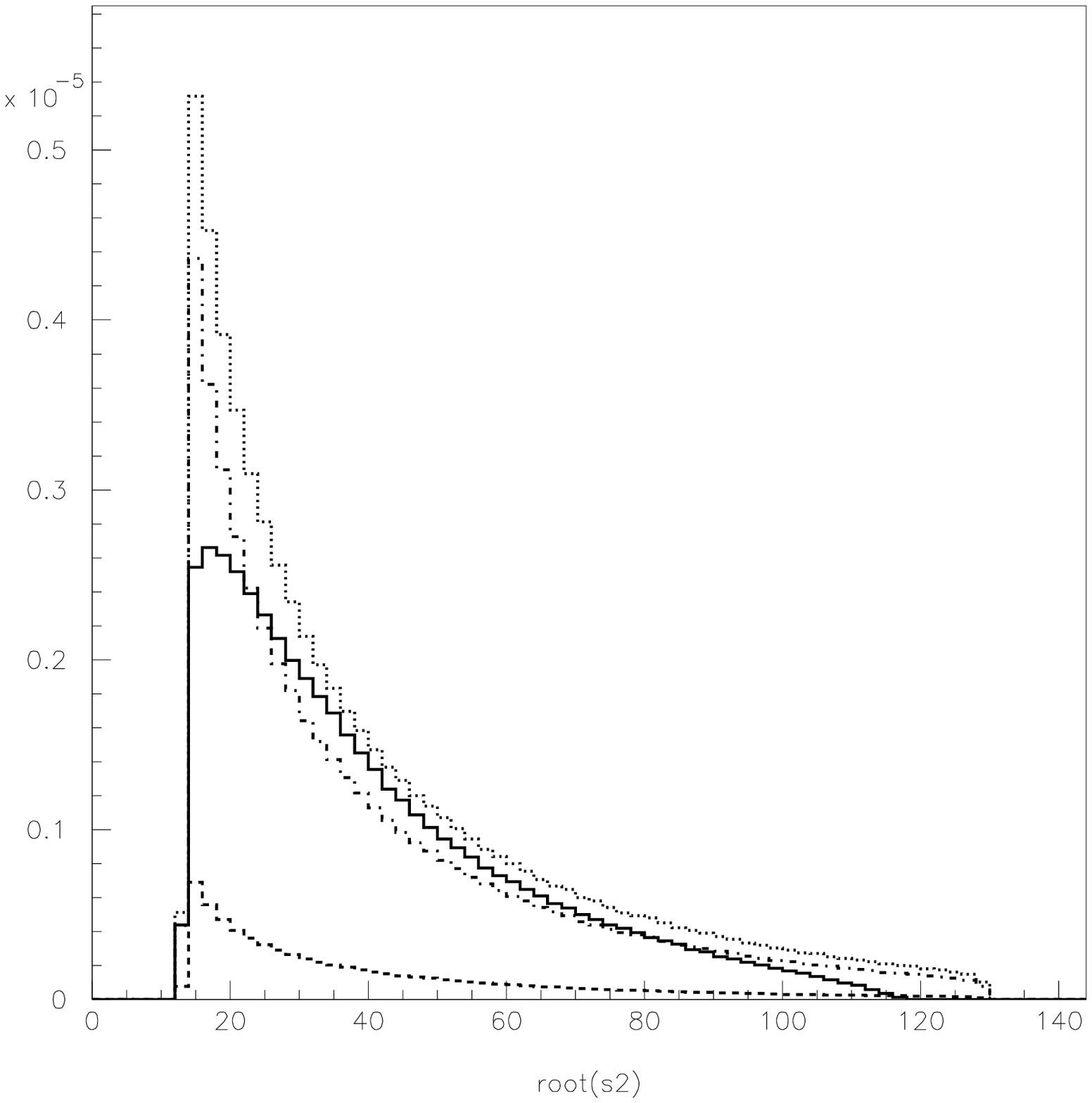,width=0.45\textwidth}
 \\
   \epsfig{file=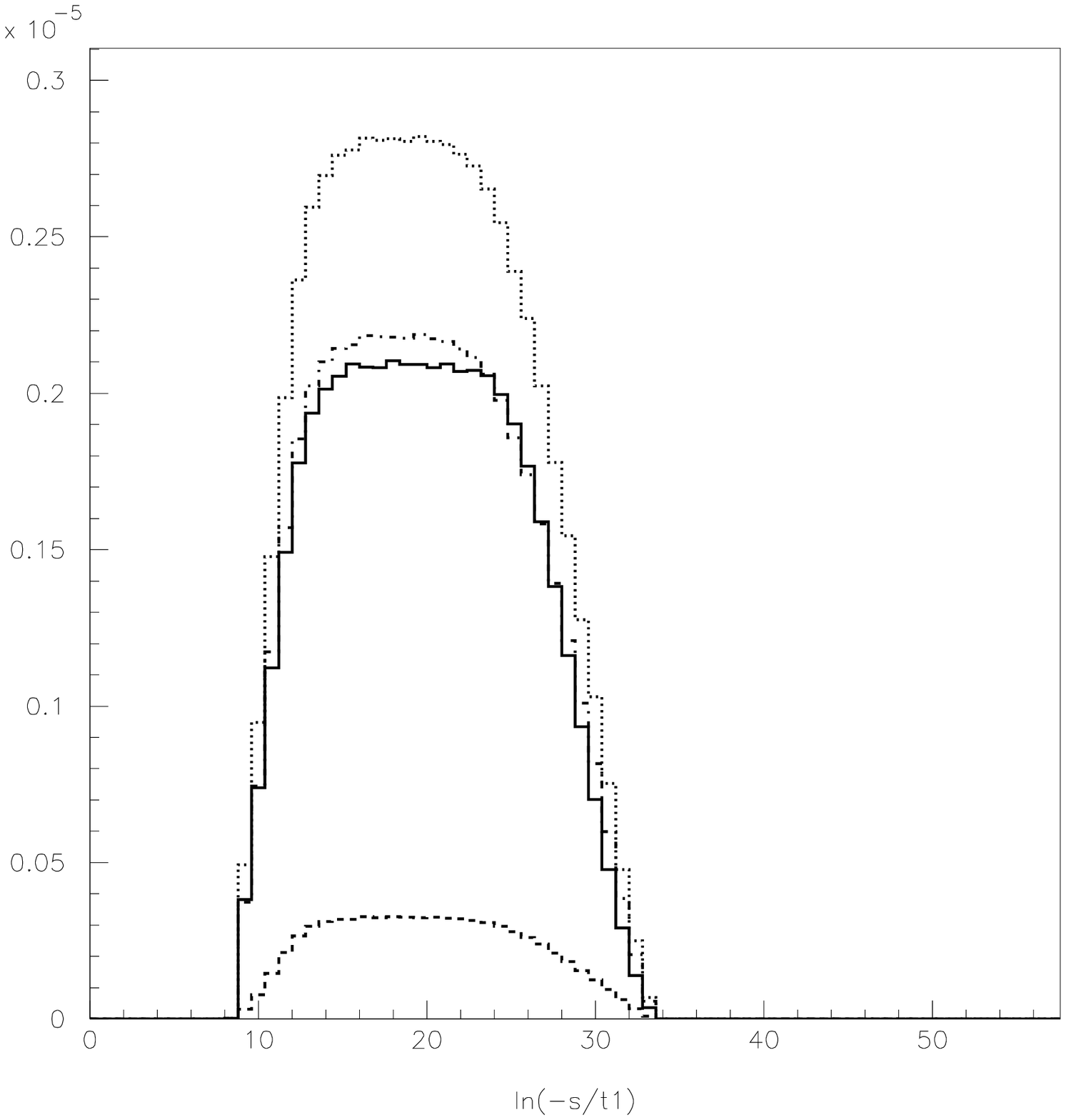,width=0.45\textwidth} &
   \epsfig{file=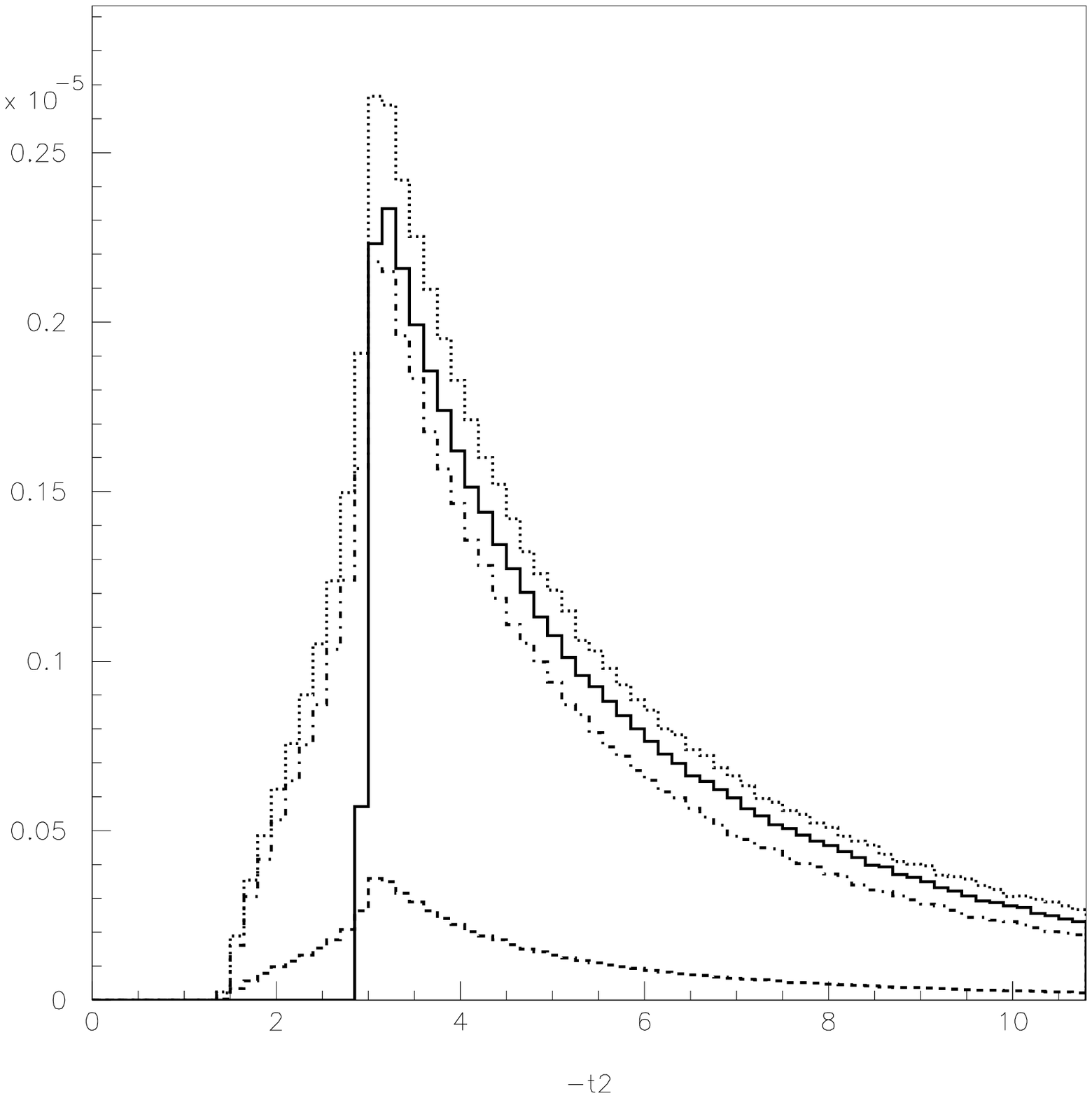,width=0.45\textwidth}
  \end{tabular}
\caption[]{Same as Fig.~\ref{fig:cutone}, but for the 
distributions in $\sqrt{s_i}$, $\ln(-s/t_1)$, and $t_2$.
\label{fig:cuttwo}}
 \end{center}
\end{figure}

\begin{figure}
 \begin{center}
\begin{tabular}[t]{c}
   \epsfig{file=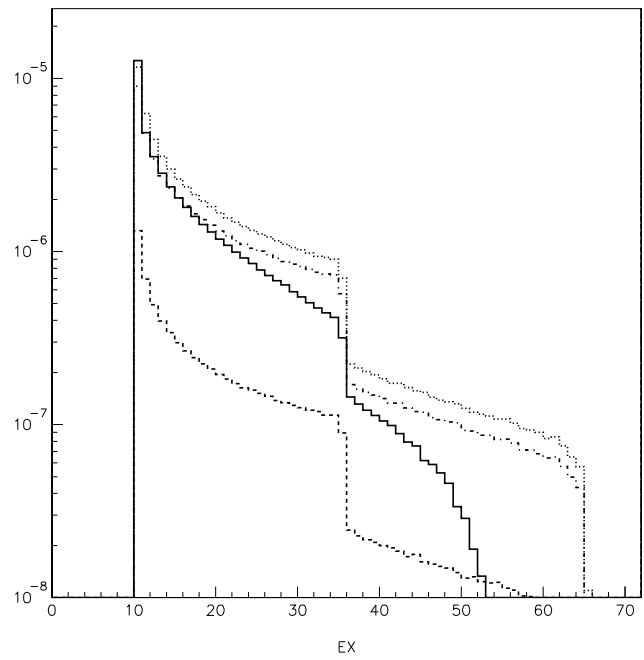,height=0.40\textheight} 
 \\
   \epsfig{file=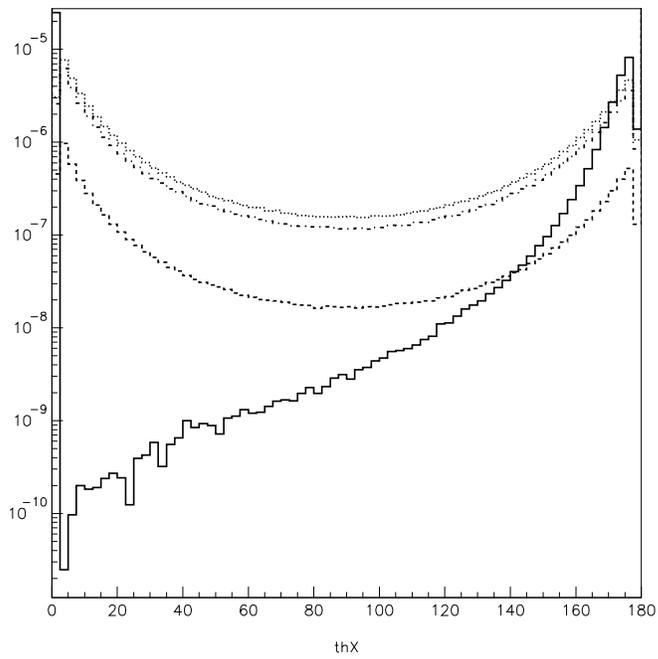,height=0.40\textheight} 
  \end{tabular}
\caption[]{Same as Fig.~\ref{fig:cutone}, but for the 
distributions in $E_X$ and $\theta_X$.
\label{fig:cutthree}}
 \end{center}
\end{figure}

Next we study the (integrated) total hadronic cross section. 
Figures~\ref{fig:nocuts} and~\ref{fig:cutone}--\ref{fig:cutthree} 
compare different ans\"atze for the $Q_i^2$ behaviour of the various 
cross sections for transverse and longitudinal photons. The 
results of the two models of generalized-vector-meson-dominance type 
(GVMD (\ref{GVMDform}) and VMDc (\ref{VMDcform}), 
dash-dotted and dotted histograms, respectively) 
are hardly distinguishable in the no-tag case, but may deviate by more than
$20\%$ in a single-tag case. In the contrast, the different $Q_i^2$ behaviour 
of a simple $\rho$-pole (dashed histograms) 
shows up already in the no-tag mode. Note that this model includes scalar 
photon contributions, but does not possess an $1/Q^2$ ``continuum'' term
for transverse photons. These differences imply that effects of non-zero 
$Q_i^2$ values must not be neglected for a precision 
measurement of $\sigma_{\gamma\gamma}(W^2)$. 

During the course of the LEP2 workshop, sophisticated programs to generate 
the full (differential) hadronic final state in two-photon collisions 
have been developed \cite{gagaMC}. The description of hadronic physics 
with one (or both) photons off-shell by virtualities $Q_i^2 \ll W^2$ 
is still premature. Indeed, existing
programs are thus far for {\em real} photons and hence use, in one
way or another, the equivalent-photon approximation (EPA) to embed the
two-photon reactions in the $\e^+\e^-$ environment. 
It is hence 
indispensable to check the uncertainties associated with the EPA. 
Hadronic physics is under much better theoretical control 
for deep-inelastic scattering, i.e.\ the setup of one almost real 
photon probed by the other that is off-shell by an amount $Q^2$ of the 
order of $W^2$. Corresponding event generators exist \cite{gagaMC} but
also in this case it is desirable to check the equivalent-photon 
treatment of the probed photon. 

An improved EPA has recently been suggested in \cite{GAS}. In essence, 
the  prescription consists in neglecting $Q_i^2$ w.r.t.\ $W^2$ 
in the kinematics but to keep the full $Q_i^2$ dependence in the 
$\gamma^\star\gamma^\star$ structure functions. In addition, non-logarithmic 
terms proportional to $m^2/Q_i^2$ in the luminosity functions are kept
as well. The study \cite{GAS} shows that this improved EPA works rather
well for the {\em integrated} $\e^+\e^-$ cross section. 
In Fig.~\ref{fig:nocuts} we show that this EPA (solid compared to
dash-dotted histograms) works well also for differential distributions, 
with the exception of the polar-angle distribution of the hadronic system 
at large angles, where it can, in fact, fail by more than an order of 
magnitude! (There, of course, the cross section is down by several orders.)

The EPA describes also rather well the dynamics of the scattered 
electrons in the single-tag mode except in the tails of the distributions
(Fig.~\ref{fig:cutone}). The same holds for the distributions in the 
photon virtualities, see Fig.~\ref{fig:cuttwo}. Sizeable differences  
do, however, show up (Fig.~\ref{fig:cuttwo})
in the distributions of the subsystem invariant masses
$\sqrt{s_i}$. These then lead to the wrong shapes for the energy and 
momentum distributions of the hadronic system shown in 
Fig.~\ref{fig:cutthree}. The EPA should, therefore, not be used for 
single-tag studies. 

\begin{figure}
 \begin{center}
%  \fbox{
\begin{tabular}[t]{c}
%   \subfigure[]{\epsfig{file=phtnvspht.eps,width=0.45\textwidth}} &
%   \subfigure[]{\epsfig{file=phitilde.eps,width=0.45\textwidth}}
   \epsfig{file=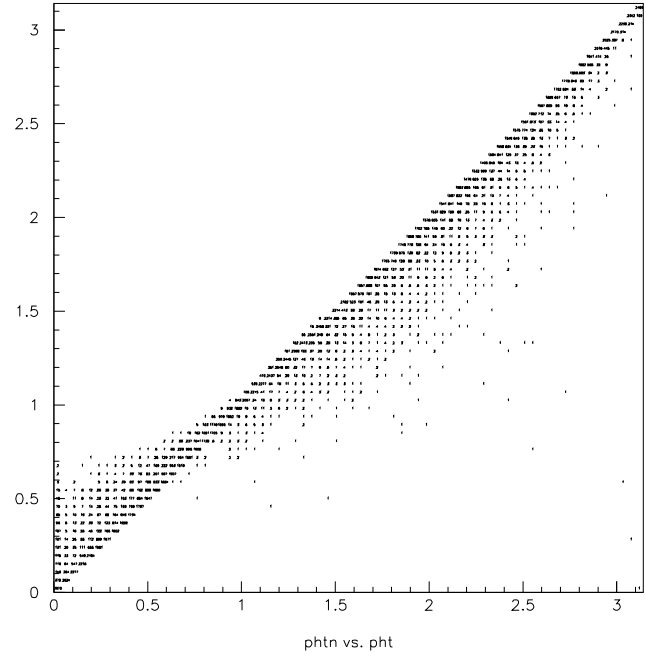,height=0.40\textheight} \\
   \epsfig{file=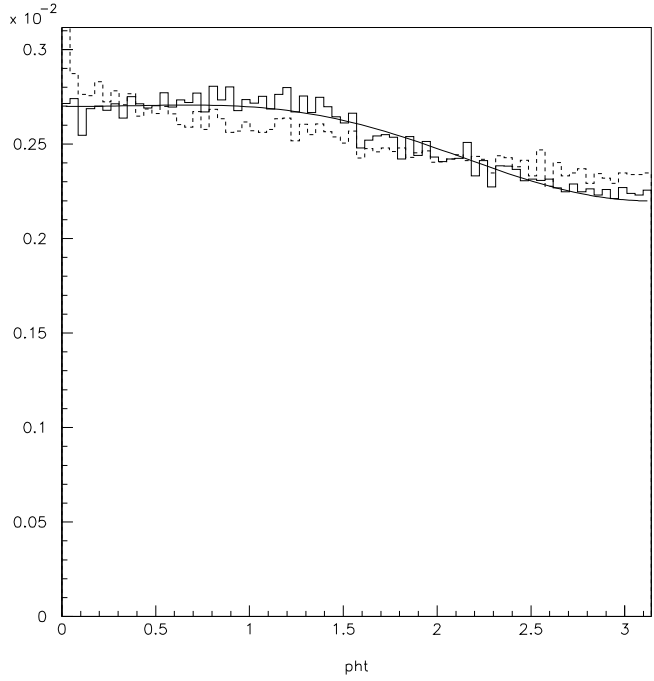,height=0.40\textheight}
  \end{tabular}
%       }
\caption[]{At the top, the correlation between $\tilde\phi_{\mrm{approx}}$ 
(\ref{phitildeapprox}), proposed in \cite{Arteaga}, and 
$\tilde{\phi}$; at the bottom, the distribution in $\tilde{\phi}$
(solid histogram) and its approximation (dashed histogram) for 
the integrated muon-pair cross section at $\sqrt{s}=130\,$GeV and 
$W=10\,$GeV. 
The cuts $1.55^\circ < \theta_i$, $5\,$GeV$ < E_1$, and
$30\,$GeV$< E_2$ have been applied.
Also shown is a fit to $\d\sigma/ \d\tilde{\phi}$ of the form
$1 + A_1 \cos\tilde\phi + A_2 \cos 2\tilde\phi$.
\label{fig:phtnvspht}}
 \end{center}
\end{figure}

Finally we study the prospects of a determination of additional structure 
functions besides $F_2$. One such possibility was outlined in \cite{LEP2}, 
namely the study of the azimuthal dependence in the $\gamma\gamma$ c.m.s.\ 
between the plane of the scattered (tagged) electron and the plane 
spanned by the beam axis and the outgoing muon or jet.
Here we propose to study the azimuthal angle $\tilde\phi$ between the two 
electron scattering planes, again in the $\gamma\gamma$ c.m.s. Although 
such a study requires a double-tag setup, the event rates need not be 
small, since one can fully integrate out the hadronic system but for its 
invariant mass $W$. In order to demonstrate the sensitivity of such
a measurement we show, as a preparatory exercise, the 
$\tilde\phi$ distribution for muon-pair production in 
Fig.~\ref{fig:phtnvspht}. Fitting to the functional form 
\begin{equation}
 \frac{\d\sigma}{\d \tilde\phi} \propto 1 + A_1\, \cos\tilde\phi
  + A_2\, \cos 2\, \tilde\phi
\ ,
\label{phitildeform}
\end{equation}
we find 
\begin{equation}
A_1 = 0.098 \qquad , \qquad A_2 = -0.028
\ .
\label{fitresult}
\end{equation}
Let us emphasize that the 
selected tagging ranges have in no way been optimized for such a study. 
Nonetheless, given the magnitudes of $A_i$, a measurement appears feasible.

All but one \cite{vermaseren} event generators for two-photon physics 
use the azimuthal angle $\phi$ between the two scattering 
planes in the {\em laboratory} frame as one of the integration variables. 
In fact, $\phi$ appears as a trivial variable in these programs. 
None of these up to now provides the calculation of $\tilde\phi$. 
An expression for $\tilde\phi$ in terms of $t_i$, $\phi$, and two other
invariants is given in \cite{Budnev} (see (\ref{phibudnev}) below) and, 
in principle, is available in TWOGAM \cite{twogam}. However, the factor
$\sqrt{t_1 t_2}$ appears explicitly in the denominator of $\cos\tilde\phi$ 
but not in its numerator. Hence, at small values of $-t_i$ this factor
will be the result of the cancellation of several much larger terms, 
rendering this expression for $\cos\tilde\phi$ numerically very unstable. 
(Recall that $|t_i|_{\min} \sim m^2(W/\sqrt{s})^4 \ll m^2$, while the 
numerator contains terms of order $s$.) In contrast, we use the 
numerically stable expression given in (\ref{phistable})\footnote{%
This form of $\tilde\phi$ could, with only minor modifications, 
be implemented in \cite{vermaseren}.}. 

An approximation for $\tilde\phi$ in terms of $\phi$ is proposed in 
\cite{Arteaga}:
\begin{equation}
 \cos\tilde\phi_{\mrm{approx}} = \cos\phi + \sin^2\phi\, 
  \frac{ Q_1 \, Q_2\, (2\, s - s_1 - s_2)}
       {(W^2 - t_1 - t_2) \sqrt{(s-s_1) (s-s_2)}}
\ .
\label{phitildeapprox}
\end{equation}
Indeed, the correlation between $\tilde\phi$ and its approximation is very
high in the no-tag case, where, however, the dependence on $\tilde\phi$ 
is almost trivial (i.e.\ flat). Figure~\ref{fig:phtnvspht} exhibits that
there is still a correlation for a double-tag mode, but formula 
(\ref{phitildeapprox}) fails to reproduce the correct $\tilde\phi$ 
dependence: a fit to (\ref{phitildeform}) yields $A_1 = 0.084$ and 
$A_2 = 0.017$, quite different from (\ref{fitresult}).

\section{Notation and cross sections}
\label{sec:notation}
Consider the reaction 
\begin{equation}
  \e^+(p_a) + \e^-(p_b) \rightarrow \e^+(p_1) + X(p_X) + \e^-(p_2)
\label{eereact}
\end{equation}
proceeding through the two-photon process
\begin{equation}
  \gamma(q_1) + \gamma(q_2) \rightarrow X(p_X)
\label{ggreact}
\ .
\end{equation}
The cross section for (\ref{eereact}) depends on six invariants, which we
choose to be the $\e^+\e^-$ c.m.\ energy $\sqrt{s}$, the $\gamma\gamma$ 
c.m.\ (or hadronic) energy $W$, the photon virtualities $Q_i$, and the
subsystem invariant masses $\sqrt{s_i}$:
\begin{eqnarray}
  \,s = (p_a + p_b)^2 & &  \quad , \quad W^2 = p_X^2
%    \quad , \quad  p_a^2 = p_b^2 = p_1^2 = p_2^2 = m^2
\ ,
\nonumber\\
  s_1 = (p_1 + p_X)^2 = (p_a + q_2)^2
& & \quad , \quad - Q_1^2 = t_1 = q_1^2 \equiv (p_a - p_1)^2 
\ ,
\nonumber\\
  s_2 = (p_2 + p_X)^2 = (p_b + q_1)^2
& & \quad , \quad - Q_2^2 = t_2 = q_2^2 \equiv (p_b - p_2)^2
\ .
\label{invdef}
\end{eqnarray}
We find it convenient to introduce also the dependent variables:
\begin{eqnarray}
  u_2 = s_1 - m^2 - t_2 & & \quad , \quad 
     \nu = \frac{1}{2}\, \left(W^2 - t_1 - t_2 \right)
\ ,
\nonumber\\
  u_1 = s_2 - m^2 - t_1 & & \quad , \quad 
      K = \frac{1}{2\, W}\, \sqrt{ \lambda(W^2,t_1,t_2) }
  = \frac{1}{W}\, \sqrt{ \nu^2 - t_1\, t_2 }
\ ,
\nonumber\\
  \, \beta = \sqrt{1 - \frac{4 m^2}{s}} & & \quad , \quad 
  y_i = \sqrt{1 - \frac{4 m^2}{t_i}} 
\ ,
\label{notation}
\end{eqnarray}
where $\lambda(x,y,z) = (x-y-z)^2-4yz$ and
$m$ denotes the electron mass. Note that $K$ is the photon 
three-momentum in the $\gamma\gamma$ c.m.s.
In terms of these variables the $\e^+\e^-$ cross section at fixed values 
of $\sqrt{s}$ and $\tau = W^2/s$ is given by:
\begin{equation}
 \frac{\d \sigma[ \e^+ \e^- \rightarrow \e^+ + \e^- + X]}{\d \tau}
  = \frac{\alpha^2\, K\,W}{2\, \pi^4\, Q_1^2\, Q_2^2 \beta}\, 
  \d R_3\, 
  \Sigma(W^2,Q_1^2,Q_2^2,s_1,s_2,\tilde{\phi};s,m^2)
\ ,
\label{crossee}
\end{equation}
where $R_3$ is the phase space for (\ref{eereact}). 

The hadronic physics is fully encoded in five structure functions. 
Three of these can be expressed through the cross sections 
$\sigma_{ab}$ for scalar ($a,b=S$) and transverse photons ($a,b=T$) 
($\sigma_{ST} = \sigma_{TS}(q_1 \leftrightarrow q_2)$). 
The other two structure functions $\tau_{TT}$ and $\tau_{TS}$
correspond to transitions with spin-flip
for each of the photons with total helicity conservation. 
%We emphasize that the hadronic physics is fully encoded in these structure 
%functions while the connection with the measured $\e^+\e^-$ cross section
%is a pure matter of QED. This connection is most transparent \footnote{%
%See, for example, \cite{Budnev} where also explicit expressions for the
%density matrices $\rho_i^{ab}$ of virtual photons can be found.}
Introducing $\tilde{\phi}$, the angle between the scattering 
planes of the colliding $\e^+$ and $\e^-$ in the {\em photon} c.m.s., 
these structure functions enter the cross sections as:
\begin{eqnarray}
  \Sigma & = & 2\, \rho_1^{++}\, 2\, \rho_2^{++}\, \sigma_{TT}
        + 2\, \rho_1^{++}\, \rho_2^{00}\, \sigma_{TS}
        + \rho_1^{00}\, 2\, \rho_2^{++}\, \sigma_{ST}
        + \rho_1^{00}\, \rho_2^{00}\, \sigma_{SS}
\nonumber\\
  & &      +~ 2\, |\rho_1^{+-}\, \rho_2^{+-}|\, \tau_{TT}\, 
         \cos2\, \tilde{\phi}
        - 8\, |\rho_1^{+0}\, \rho_2^{+0}|\, \tau_{TS}\,
          \cos\tilde{\phi}
\ .
\label{Sigmadef}
\end{eqnarray}
The density matrices of the virtual photons in the $\gamma\gamma$-helicity
basis are given by
\begin{eqnarray}
  2\, \rho_1^{++} & = & \frac{\left( u_2 - \nu\right)^2}{K^2\, W^2}
      + 1 + \frac{4\, m^2}{t_1}
\nonumber\\
   \rho_1^{00} & = & \frac{\left( u_2 - \nu\right)^2}{K^2\, W^2}
      - 1 
\nonumber\\
   |\rho_1^{+-}| & = &  \rho_1^{++} - 1
\nonumber\\
   |\rho_1^{+0}| & = &  \sqrt{\left(\rho_1^{00} + 1\right)|\rho_1^{+-}|}
  = \frac{u_2 - \nu}{K\, W}\, \sqrt{ \rho_{1}^{++} - 1}
\label{rhoidef}
\end{eqnarray}
with analogous formulas for photon $2$.

A few remarks about the numerical stability of the $\tilde\phi$-dependent 
terms are in order. Thus far, these terms are implemented solely in the 
TWOGAM \cite{twogam} event generator, using the formulas quoted in 
\cite{Budnev}. Given in \cite{Budnev} and coded
in \cite{twogam} are the products 
$X_2 = 2\, |\rho_1^{+-}\, \rho_2^{+-}|\,  \cos2\, \tilde{\phi}$ and 
$X_1 = 8\, |\rho_1^{+0}\, \rho_2^{+0}|\,  \cos\tilde{\phi}$ in terms
of invariants. Now, the expressions for $X_i$ contain explicit factors 
of $t_1 t_2$ ($X_2$) and $\sqrt{t_1 t_2}$ ($X_1$) in the denominators 
but not in the numerators. Clearly,  the evaluation of $X_i$ becomes 
unstable for small values of $|t_i|$. On the other hand, the factors 
multiplying $\cos\tilde\phi$ and $\cos 2\tilde\phi$ in $X_i$ approach 
perfectly stable expressions in the limit $m^2/W^2 \rightarrow 0$ and
$t_i/W^2 \rightarrow 0$:
\begin{eqnarray}
   |\rho_1^{+-}| & \rightarrow & \frac{2}{x_1^2}\, (1 - x_1)
  + \frac{2\, m^2}{t_1}
\nonumber\\
   |\rho_1^{+0}| & \rightarrow &  \frac{2 - x_1}{x_1}\, 
            \sqrt{ |\rho_{1}^{+-}| }
\ ,
\end{eqnarray}
where $x_i = W^2/s_i \approx s_k/s$ ($i \neq k$). Hence a numerically
stable evaluation of $\tilde\phi$ guarantees a correct evaluation of
the $\tilde\phi$-dependent terms.

The structure functions $\sigma_{ab}$ and $\tau_{ab}$ for lepton-pair 
production are well quoted in the literature; the formulas of 
\cite{Budnev} are implemented in the program. Much less is known 
about the structure functions for hadronic processes. 
Since we are not aware of a model for $\tau_{ab}$, the current version 
of the program assumes
\begin{equation}
  \tau_{TT} = 0 = \tau_{TS}
\ .
\label{taudef}
\end{equation}
The cross sections $\sigma_{ab}$ are uncertain at small 
values of $Q_i$. Three models for $\sigma_{ab}$ are provided, all
based upon the assumption 
\begin{equation}
  \sigma_{ab}(W^2,Q_i^2) = h_a(Q_1^2)\, h_b(Q_2^2)\, 
\sigma_{\gamma\gamma}(W^2) 
\ ,
\label{sigapprox}
\end{equation}
which is valid for $Q_i^2 \ll W^2$, which is justified in most applications. 
Note the cross section for the scattering of two real photons 
$\sigma_{\gamma\gamma}(W^2)$ that enters as a multiplicative factor 
in (\ref{sigapprox}).

The three models are defined as follows. 
The first one is based upon a parametrization \cite{Bezrukov}
of the $\gamma^* \p$ cross section calculated in a model of 
generalized vector-meson dominance (GVMD):
\begin{eqnarray}
  h_T(Q^2) & = & r\, P_1^{-2}(Q^2) + (1-r)\, P_2^{-1}(Q^2) 
\nonumber\\
  h_S(Q^2) & = & \xi\, \left\{ r\, \frac{Q^2}{m_1^2}\, P_1^{-2}(Q^2)
       + (1-r)\, \left[ \frac{m_2^2}{Q^2} \, \ln P_2(Q^2) 
       -  P_2^{-1}(Q^2) \right] \right\}
\nonumber\\
  P_i(Q^2) & = & 1 + \frac{Q^2}{m_i^2}
\ ,
\label{GVMDform}
\end{eqnarray}
where we take $\xi=1/4$, $r=3/4$, $m_1^2 = 0.54\,$GeV$^2$ 
and $m_2^2 = 1.8\,$GeV$^2$.

The second model \cite{Sakurai} adds a continuum contribution to simple 
(diagonal, three-mesons only) vector--meson dominance (VMDc):
\begin{eqnarray}
  h_T(Q^2) & = & \sum_{V=\rho,\omega,\rho}\, r_V\, 
      \left(\frac{m_V^2}{m_V^2 + Q^2}\right)^2
  + r_c\, \frac{m_0^2}{m_0^2 + Q^2}
\nonumber\\
  h_S(Q^2) & = & \sum_{V=\rho,\omega,\rho}\, \frac{\xi\, Q^2}{m_V^2}\, 
      r_V\,  \left(\frac{m_V^2}{m_V^2 + Q^2}\right)^2
\ ,
\label{VMDcform}
\end{eqnarray}
where
$r_\rho=0.65$, $r_\omega = 0.08$, 
$r_\phi = 0 .05$, and $r_c = 1 - \sum_V r_V$. 

Since photon-virtuality effects are often estimated by using a simple 
$\rho$-pole 
only, we include also the model defined by ($\rho$-pole):
\begin{equation}
  h_T(Q^2)  =  \left(\frac{m_\rho^2}{m_\rho^2 + Q^2}\right)^2
\qquad , \qquad
  h_S(Q^2)  =  \frac{\xi\, Q^2}{m_\rho^2}\, 
      \left(\frac{m_\rho^2}{m_\rho^2 + Q^2}\right)^2
\ .
\label{rhoform}
\end{equation}
Since the program is meant to be used at fixed $W$, we take
\begin{equation}
 \sigma_{\gamma\gamma}(W) = 1
\ .
\end{equation}

Finally we give the relation between the cross section at fixed values 
of $\tau$ and $Q_2^2$ and the usual form used in deep-inelastic scattering:
\begin{equation}
  \frac{\d \sigma}{\d \tau\, \d t_2} = \frac{x^2\, s}{Q_2^2}\, 
  \frac{\d \sigma}{\d x\, \d Q_2^2}
\ ,
\label{DIScross}
\end{equation}
where $x$ is the Bjorken-$x$ variable defined by
\begin{equation}
  x = \frac{Q_2^2}{2\, q_1\cdot q_2} = \frac{Q_2^2}{W^2 + Q_2^2 + Q_1^2}
\ .
\label{xBdef}
\end{equation}

\section{Phase space}
\label{sec:phase}
The phase space can be expressed in terms of four invariants: 
\begin{eqnarray}
 \d R_3 & \equiv & \prod_{i=1,2,X}\, \int\, \frac{\d^3 p_i}{2\, E_i}\,
  \delta^4 \left( p_a + p_b - \sum_i\, p_i \right)
\nonumber\\
  & = & 
 \frac{1}{16\,\beta\,s}\, 
   \int\,\d t_2\, \d t_1\, \d s_1 \, \d s_2\, 
            \frac{\pi}{\sqrt{ - \Delta_4}}
\ ,
\end{eqnarray}
where $\Delta_4$ is the $4\times 4$ symmetric Gram determinant of any 
four independent vectors formed out of $p_a$, $p_b$, $p_1$, $p_X$, $p_2$. 
The physical region in $t_2$, $t_1$, $s_1$, $s_2$ for fixed $s$ satisfies
$\Delta_4 \leq 0$. Since $\Delta_4$ is a quadratic polynomial in any of 
its arguments, the boundary of the physical region, $\Delta_4 = 0$, 
is a quadratic equation and has two solutions. Picking $s_2$ as the 
innermost integration variable, the explicit evaluation of $\Delta_4$ yields
\begin{equation}
 16\, \Delta_4  =  a\,s_2^2 + b\,s_2 + c 
      = a\, (s_2 - s_{2+})\, (s_2 - s_{2-})
\ ,
\end{equation}
where
\begin{eqnarray}
a & = &
  \lambda(s_1, t_2, m^2)
%t_2^{2}+s_1^{2}-2\,t_2\,s_1
%-2\,{m}^{2}s_1-2\,{m}^{2}t_2+{m}^{4}
\nonumber\\
b & = & -2\,s\ {m}^{2}t_1-2\,{m}^{2}s_1^{2}+8\,
t_2\,{m}^{4}-2\,{m}^{2}t_2^{2}-2\,s\ s_1\,{W}^{2}
+2\,{m}^{2}s\ {W}^{2}+2\,t_1\,s\ s_1+2\,s\ t_2\,s_1
\nonumber\\ & &~
+4\,{m}^{2}s_1\,{W}^{2}+4\,{m}^{4}s_1+2\,t_1\,
t_2\,s-2\,t_2\,{m}^{2}t_1-2\,t_2^{2}s
-2\,{m}^{2}t_2\,s+2\,t_1\,t_2\,s_1
\nonumber\\ & &~
-4\,{m}^{4}{W}^{2}
-2\,t_1\,s_1^{2}+2\,s\ t_2\,{W}^{2}-2\,{m}^{6}
+2\,{m}^{4}t_1
\nonumber\\
c & = & -2\,s\ {m}^{4}{W}^{2}-2\,t_1^{2}{m}^{2}s_1
-2\,t_1\,t_2\,{s}^{2}+2\,s\ t_1\,t_2\,s_1-2\,
s\ t_1^{2}s_1+t_1^{2}{s}^{2}+t_1^{2}s_1^{2}
+t_2^{2}{s}^{2}
\nonumber\\ & &~
+{m}^{4}s_1^{2}+{m}^{4}t_1^{2}-6
\,{m}^{6}t_1-2\,{m}^{6}s_1-4\,{m}^{4}s_1\,{W}^{2}+2\,
{m}^{4}t_2\,s+2\,{m}^{4}t_2\,s_1+8\,{m}^{4}t_1\,s_1
\nonumber\\ & &~
-2\,{s}^{2}t_2\,{W}^{2}-2\,t_1\,{s}^{2}{W}^{2}-2\,{m}
^{2}t_1\,s_1^{2}+{m}^{8}-2\,{m}^{2}s\ t_2\,s_1+4\,
{m}^{6}{W}^{2}+{m}^{4}t_2^{2}
\nonumber\\ & &~
+4\,{m}^{2}t_1\,t_2\,s-2
\,{m}^{2}t_1\,t_2\,s_1-6\,{m}^{6}t_2+{s}^{2}{W}^{4
}+6\,{m}^{2}s\ t_2\,{W}^{2}-4\,s\ {m}^{2}{W}^{4}
\nonumber\\ & &~
-2\,s\ {m}^{2}t_1^{2}+2\,s\ t_1\,{m}^{4}-2\,s\ {m}^{2}t_2^{2}
+2\,t_1\,t_2\,{m}^{4}+2\,s\ {m}^{2}s_1\,{W}^{2}-4\,s\ t_1\,t_2\,
{W}^{2}
\nonumber\\ & &~
-2\,s\ t_1\,{m}^{2}s_1+6\,s\ t_1\,{m}^{2}{W}^{2}+2\,
s\ t_1\,s_1\,{W}^{2}
\ .
\label{abcdef}
\end{eqnarray}
A numerical stable form for the $s_2$ limits is
\begin{eqnarray}
  s_{2+} & = & \frac{-b + \sqrt{\Delta}}{2\, a}
\nonumber\\
  s_{2-} & = & \frac{c}{a\, s_{2+}}
\ ,
\label{s2maxmin}
\end{eqnarray}
where
%\begin{eqnarray}
%\Delta & \equiv & b^2 - 4\, a\, c
%\nonumber\\
%& = & 16\,\left (-2\,t_1\,{m}^{2}s_1-t_2\,{m}
%^{2}t_1+{m}^{4}t_1-{m}^{2}{W}^{2}t_1+{m}^{2}t_2^
%{2}+t_2\,{W}^{2}t_1-t_1\,s_1\,{W}^{2}
%\right.
%\nonumber\\ & &~
%\left.
%-2\,{m}^{2}
%t_2\,{W}^{2}+{m}^{2}{W}^{4}+t_1\,s_1^{2}+t_1^{2}
%s_1-t_1\,t_2\,s_1\right )
%\nonumber\\ & &~
%\left ({m}^{2}s_1
%^{2}-2\,{m}^{4}s_1-s\ t_2\,s_1-3\,{m}^{2}t_2\,s\ +{m}^
%{6}+t_2\,{s}^{2}+t_2^{2}s\right )
%\label{Deltadef}
%\end{eqnarray}
$\Delta = b^2 - 4\, a\, c$ is given below in a numerically stable form, 
in (\ref{Deltadef}).

In order to remove the singularity due to $(-\Delta_4)^{-1/2}$
(in the limit $|t_i|$, $m^2 \ll s_i$, $W^2$, the $s_2$ integration 
degenerates to an integration over the $\delta$-function
$\delta(s_2 - s\, W^2/s_1)$), 
it is advisable to change variable from $s_2$ to $x_4$, 
$0 \leq x_4 \leq 1$:
\begin{eqnarray}
      s_2  & = &  \frac{1}{2\, a}\, 
       \left\{ - b - \sqrt{\Delta}\, 
           \cos\left( x_4\, \pi\right) \right\}
\nonumber\\
  \int_{s_{2-}}^{s_{2+}}\, \frac{ \d s_2}{\sqrt{-\Delta_4}}
  & = & \frac{4\, \pi}{\sqrt{a}}\, \int_0^1\, \d x_4
\ .
\label{s2vartrafo}
\end{eqnarray}
For later use we also need a numerically stable form of the Gram 
determinant, which reads 
\begin{equation}
  16\, \Delta_4 = - \frac{\Delta\, \sin^2\left( x_4 \, \pi\right)}{4\, a}
\ .
\end{equation}

The $s_1$-integration limits follow from the requirement $\Delta > 0$. 
They are most easily derived when realizing that the discriminant $\Delta$
is given as the product of two $3\times 3$ symmetric Gram determinants or, 
equivalently, the product of two kinematic $G$ functions
\begin{equation}
  \frac{1}{4}\, \Delta = 4\, G_3\, G_4 = 64\, D_3\, D_4
\ ,
\label{Deltadef}
\end{equation}
where
\begin{eqnarray}
  -4\, D_3  \equiv  -4\, \Delta_3(p_a,p_b,q_2)
 & = & G(s, t_2, s_1, m^2, m^2, m^2) \equiv G_3
\nonumber\\
  -4\, D_4  \equiv  -4\, \Delta_3(p_a,q_1,q_2)
 & = & G(t_1, s_1, t_2, m^2, m^2, W^2) \equiv G_4
\ .
\label{Didef}
\end{eqnarray}
Since any $3\times 3$ Gram determinant $\Delta_3$ satisfies 
$\Delta_3 \geq 0$, the physical region is that where
both $G_3$ and $G_4$ are simultaneously negative. Solving $G_i$ for $s_1$
\begin{eqnarray}
  G_3 & = & m^2\, (s_1 - s_{11+})\, (s_1 - s_{11-})
\nonumber\\
 & = & {m}^{2}s_1^{2}-2\,{m}^{4}s_1-s\ t_2\,s_1-3\,{m}^{2}t_2\,s\ 
   +{m}^{6}+t_2\,{s}^{2}+t_2^{2}s
\nonumber\\
  G_4 & = & t_1\, (s_1 - s_{12+})\, (s_1 - s_{12-})
\nonumber\\
  & = & -2\,t_1\,{m}^{2}s_1-t_2\,{m}^{2}t_1+{m}^{4}t_1-{m}^{2}{W}^{2}t_1
  +{m}^{2}t_2^{2}+t_2\,{W}^{2}t_1-t_1\,s_1\,{W}^{2}
\nonumber\\ & &~
-2\,{m}^{2}t_2\,{W}^{2}+{m}^{2}{W}^{4}+t_1\,s_1^{2}+t_1^{2}
s_1-t_1\,t_2\,s_1
\label{Gidef}
\end{eqnarray}
we find
\begin{eqnarray}
s_{11\pm}& = & {\frac {t_2\,S+2\,{m}^{4}\pm \sqrt {\lambda(S,{m}^{2}
,{m}^{2})\lambda(t_2,{m}^{2},{m}^{2})}}{2\,{m}^{2}}}
\nonumber\\
s_{12\pm} & = & {\frac {t_2}{2}}+{m}^{2}+{\frac {{W}^{2}}{2}}-{
\frac {t_1}{2}} \pm {\frac {\sqrt {\lambda(t_1,t_2,{W}^{2})
\lambda(t_1,{m}^{2},{m}^{2})}}{2\,t_1}}
\nonumber\\
s_{11+} s_{11-} & = & {\frac {t_2\,S\left (-3\,{m}^{2}+S+t_2
\right )}{{m}^{2}}}+{m}^{4}
\nonumber\\
s_{12+} s_{12-} & = & 
%-\left (m-W\right )\left (m+W\right )
  \left( W^2 - m^2 \right)\,
\left (-{m}^{2}
+t_2\right )+{\frac {{m}^{2}\left ({W}^{2}-t_2\right )^{2}}
{t_1}}
\ .
\label{sijdef}
\end{eqnarray}
Note that $s_{12+} \leq s_{12-}$. Since $G_3$ is always negative between
its two roots, the range of integration over $s_1$ is
$s_{12-} \leq s_1 \leq s_{11+}$. Numerically it is more advantageous
to calculate the limits as
\begin{eqnarray}
      s_{1\min} & = & s_{12-} = m^2 + \frac{1}{2}\, 
  \left(W^2-t_1+t_2+y_1\ \sqrt{ \lambda(W^2,t_1,t_2)} \right)
\nonumber\\
      s_{1\max} & = &  s_{11+} = 
          m^2 + \frac{2\ (s+t_2-4\ m^2)}{1+\beta\ y_2}
\ .
\label{s1minmax}
\end{eqnarray}
%where
%\begin{equation}
%      y_1  =  \sqrt{1 - 4\ m^2/t_1 }
%\label{y1def}
%\end{equation}

The dominant behaviour of the $s_1$ integration is given by
the factor $\lambda^{-1/2}(s_1,t_2,m^2)$, see (\ref{s2vartrafo}).
(In the limit $t_2$, $m^2 \ll s_1$, this becomes $\d s_1/s_1$ integration.)
This factor can be transformed away by the variable transformation 
from $s_1$ to $x_3$, $0 \leq x_3 \leq 1$,
\begin{eqnarray}
      s_1   & = & X_{1}/2 + m^2 + t_2 + 2\, m^2\, t_2/X_{1}
\nonumber\\
    X_{1}   & = & (\nu + K\, W)\, (1+y_1)\, \exp{(\delta_1\, x_3)}
\nonumber\\
   \delta_1 & = & \ln\frac{s\, (1+\beta)^2}
                          {(\nu+K\, W)\, (1+y_1)\, (1+y_2)}
%\nonumber\\
%     \nu    & = & \frac{1}{2}\, \left(W^2 - t_1 - t_2 \right)
%\nonumber\\
%      K     & = &  \frac{1}{2\, W}\, \sqrt{ \lambda(W^2,t_1,t_2) }
%  = \frac{1}{W}\, \sqrt{ \nu^2 - t_1\, t_2 }
\ ,
\label{s1vartrafo}
\end{eqnarray}
such that
\begin{equation}
  \int_{s_{1\min}}^{s_{1\max}}\, \d s_1\, \frac{4\,\pi}{\sqrt{a}}
  = 4\, \pi \, \delta_1\, \int_0^1\, \d x_3
\ .
\label{x3def}
\end{equation}

The physical region in the $t_1$--$t_2$ plane is defined by the requirement
$G_i < 0$ for all $s_1$ values between the limits $(m+W)^2 \leq s_1
\leq (\sqrt{s}-m)^2$. Since for the reaction considered here
the masses of the particles involved are such that the values
$t_1 = (m_a - m_1)^2$, $t_2 = (m_b - m_2)^2$ cannot be reached and
$t_2$ is never larger than zero, the boundary curve in the $t_1$--$t_2$ 
plane is simply given by $s_{12-} = s_{11+}$. Equivalently, the 
$t_1$ limits can be found by solving $G_4=0$ with $s_1 = s_{11+}$ for $t_1$:
\begin{equation}
      t_{1\min}  =  - \frac{1}{2}\, 
         \left( \frac{b_1}{a_1} + \Delta t_1 \right)
\qquad , \qquad 
      t_{1\max}  =  \frac{ c_1}{a_1\ t_{1min} }
\ ,
\label{t1maxmin}
\end{equation}
where
\begin{eqnarray}
      \Delta t_1 & = & \frac{ \sqrt{ \Delta_1}}{a_1}
\nonumber\\
      a_1   & = & 2\ (Q+t_2+2\ m^2+W^2)
\nonumber\\
      b_1  &= & Q^2-W^4+2\ W^2\ t_2-t_2^2-8\ m^2\ t_2-8\ m^2\ W^2
\nonumber\\
      c_1 & = & 4\ m^2\, (W^2-t_2)^2
\nonumber\\
      \Delta_1 & \equiv & b_1^2 - 4\, a_1\, c_1
\nonumber\\
         & = & (Q+t_2-W^2+4\ m\ W)\, (Q+t_2-W^2-4\ m\ W)
\nonumber\\ & &~
    (Q^2-2\ Q\ t_2+2\ Q\ W^2+t_2^2+W^4-16\ m^2\ t_2-2\ W^2\ t_2)
\nonumber\\
Q  & = & \frac{1}{m^2}\, \left\{ 
  t_2\,s-{m}^{2}t_2-{m}^{2}{W}^{2}+
    \sqrt {\lambda(s,{m}^{2},{m}^{2})\lambda(t_2,{m}^{2},{m}^{2})} \right\}
\nonumber\\
       & = & \frac{ 4\ (s+t_2-4\ m^2)}{1+\beta\ y_2} - t_2 - W^2
\ .
%\nonumber\\
%      y_2  & = & \frac{1}{-t_2}\, \sqrt{\lambda(t_2,m^2,m^2)}
%     = \sqrt{1 - 4\ m^2/t_2 }
\label{t1hilf}
\end{eqnarray}

Finally, the $t_2$-integration limits follow from requiring 
$\Delta_1 \geq 0$:
\begin{equation}
  \Delta_1 = \frac{ (t_2 - t_{21+})\, (t_2 - t_{21-})}{F_1}
         \,  \frac{ (t_2 - t_{22+})\, (t_2 - t_{22-})}{F_2}
         \,  \frac{ t_2\, (t_2 - t_{23})^2}{F_3}
\ ,
\end{equation}
where
\begin{eqnarray}
F_1& = & {\frac {4\,s}{t_2\,s\beta\,y_2+t_2\,s-2\,{W}
^{2}{m}^{2}+4\,{m}^{3}W}}
\nonumber\\
F_2& = & {\frac {4\,s}{t_2\,s\beta\,y_2+t_2\,s-2\,{W}
^{2}{m}^{2}-4\,{m}^{3}W}}
\nonumber\\
F_3& = & 
- 16\, m^4 / \left\{ 
-2\,t_2\,{s}^{2}\beta\,y_2+
4\,t_2\,s\beta\,y_2\,{m}^{2}-t_2\,{s}^{2}-t_2\,{s}
^{2}{\beta}^{2}+4\,t_2\,s{m}^{2}
\right.
\nonumber\\ & &~ \left.
+4\,{s}^{2}{\beta}^{2}{m}^{2}-4\,
t_2\,{m}^{4}+16\,{m}^{6} \right\}
\nonumber\\
t_{21\pm}& = & -{\frac {2\,mW-{W}^{2}+{s}-4\,{m}^{2}\pm \beta\, \sqrt {\left 
(s-{W}^{2}\right ) 
%\left (s+4\,mW-{W}^{2}-4\,{m}^{2}\right )
\left( s - W_-^2 \right)
}}{2}}
\nonumber\\
t_{22\pm}& = & -{\frac {-2\,mW-{W}^{2}+{s}-4\,{m}^{2} \pm \beta\, \sqrt {
\left (s-{W}^{2}\right ) 
%\left (s-4\,mW-{W}^{2}-4\,{m}^{2}\right )
\left( s - W_+^2 \right)
}}{2}}
\nonumber\\
t_{23}& = & {\frac {\left (s-2\,{m}^{2}\right )^{2}}{{m}^{2}}}
\nonumber\\
 W_{\pm} & = & W \pm 2\, m
\ .
\end{eqnarray}
Equivalently, they are arrived at by solving $G_3=0$ 
with $s_1 = (m+W)^2$ for $t_2$:
\begin{eqnarray}
      t_{2\min} & = & t_{22+} = -\frac{1}{2}
   \left(s - W^2 - 2\ m\ W - 4\ m^2 + \Delta t_2 \right)
\nonumber\\
      t_{2\max} & = & t_{22-} = \frac{m^2\ W^2\ W_+^ 2}{s\ t_{2\min}}
\ ,
\label{t2maxmin}
\end{eqnarray}
where
\begin{equation}
      \Delta t_2 = \beta \sqrt{(s - W^2)(s - W_+^2)}
\ .
\label{t2hilf}
\end{equation}

The phase space finally becomes
\begin{equation}
 \d R_3 = \frac{\pi^2}{4\, \beta\, s}\,
  \int_{t_{2\min}}^{t_{2\max}}\, \d t_2\
  \int_{t_{1\min}}^{t_{1\max}}\, \d t_1\
  \delta_1(t_1,t_2)\, \int_0^1 \d x_3\, \int_0^1 \d x_4
\ .
\label{newdR3}
\end{equation}
The dominant $t_i$ behaviour is taken into account through a logarithmic
mapping, so that we end up with a cross section of the form
\begin{eqnarray}
  \frac{\d \sigma}{\d \tau}
  & = & \prod_{i=1}^{4}\,  \int \d x_i\ F(x_i)
\nonumber\\
 & \equiv & \prod_{i=1}^{4}\,  \int \d x_i\ 
  \ln\frac{t_{2\max}}{t_{2\min}}\
  \ln\frac{t_{1\max}}{t_{1\min}}\
  \ln\frac{s\, (1+\beta)^2}{(\nu + K\, W)\, (1+y_1)\, (1+y_2)}\ 
  \frac{\alpha^2\, K\, W}{8\, \pi^2\, \beta^2\, s}\ \Sigma 
\ .
\label{Fdef}
\end{eqnarray}

\section{Equivalent-photon approximation}  
\label{sec:EPA}
An approximation is arrived at by neglecting as much as possible 
the electron-mass and $t_i$ dependences in the kinematics, but keeping the 
full dependence on $W$ and $Q_i$ in the hadronic cross sections 
$\sigma_{ab}(W^2,Q_1^2,Q_2^2)$ \cite{GAS}:
\begin{eqnarray}
  \frac{\d\sigma}{\d \tau} & = & 
  \int_{W^2}^{s}\, \frac{\d s_1}{s_1}\
  \int_{t_{2a}}^{t_{2b}}\, \frac{\d t_2}{t_2}\
  \int_{t_{1a}}^{t_{1b}}\, \frac{\d t_1}{t_1}\
  \int_{W^2}^{s}\, \d s_2\  
  \delta \left( s_2 - \frac{s\, W^2}{s_1} \right)\ 
  \frac{\alpha^2\, W^2}{16\, \pi^2\, s}\, 
\nonumber\\ & &~ 
%  \Sigma_{\mrm{approx}}(W^2,Q_1^2,Q_2^2,s_1,s_2,\tilde{\phi};s,m^2)
  \left\{ 2\, \rho_{1\mrm{approx}}^{++}\, 
          2\, \rho_{2\mrm{approx}}^{++}\, \sigma_{TT}
        + 2\, \rho_{1\mrm{approx}}^{++}\, 
              \rho_{2\mrm{approx}}^{00}\, \sigma_{TS} 
\right.
\nonumber\\ & &~~ \left.
            + \rho_{1\mrm{approx}}^{00}\, 
          2\, \rho_{2\mrm{approx}}^{++}\, \sigma_{ST}
        + \rho_{1\mrm{approx}}^{00}\, 
          \rho_{2\mrm{approx}}^{00}\, \sigma_{SS} \right\}
\ .
\label{approxcross}
\end{eqnarray}
The integration limits are given by:
\begin{eqnarray}
  t_{ia} & = & - \frac{m^2\, x_i^2}{1 - x_i} - (1 - x_i)\,
     \sin^2 \frac{\theta_{i\max}}{2}
\nonumber\\
  t_{ib} & = & - \frac{m^2\, x_i^2}{1 - x_i} - (1 - x_i)\,
     \sin^2 \frac{\theta_{i\min}}{2}
\ ,
%\nonumber\\
%  x_1 & = & \frac{s_2}{s} \quad , \quad x_2 = \frac{s_1}{s}
\label{tiapproxlimits}
\end{eqnarray}
where $x_1 = s_2/s$ and $x_2 = s_1/s$. 

The approximate forms of the photon density matrices read: 
\begin{eqnarray}
  2\, \rho_{1\mrm{approx}}^{++} & = & \frac{2}{x_1^2}\, \left\{
  1 + (1-x_1)^2 - \frac{2\, m^2\, x_1^2}{Q_1^2} \right\}
\nonumber\\
   \rho_{1\mrm{approx}}^{00} & = & \frac{4}{x_1^2}\, \left(
            1 - x_1 \right)
\ .
\end{eqnarray}
\section{Momenta}
\label{sec:momenta}
Here we present the particle momenta in the laboratory frame.
The particle energies follow simply from $E_i = (p_a + p_b)\cdot p_i
/\sqrt{s}$:
\begin{eqnarray}
         E_1 & = & \frac{s + m^2 - s_2}{2\, \sqrt{s}}
\nonumber\\
         E_2 & = & \frac{s + m^2 - s_1}{2\, \sqrt{s}}
\nonumber\\
         E_X & = & \frac{s_1 + s_2 - 2\, m^2}{2\, \sqrt{s}}
\end{eqnarray}
and the moduli of the three-momenta from $P_i^2 = E_i^2 - m_i^2$. 
The polar angles $\theta_i$ with respect to the beam axis could be 
calculated from $p_b \cdot p_i = E_b\, E_i - P_b\, P_i\, \cos\theta_i$
\begin{eqnarray}
    \cos\theta_1 & = &  \frac{s- s_2+ 2\, t_1 - 3\, m^2}
                   {2\,\beta\, \sqrt{s}\, P_1}
\nonumber\\
    \cos\theta_2 & = &  \frac{s- s_1+ 2\, t_2 - 3\, m^2}
                   {2\, \beta\, \sqrt{s}\, P_2}
\nonumber\\
    \cos\theta_X & = &  - \frac{s_2-s_1+2\, (t_2-t_1)}
                   {2\, \beta\, \sqrt{s}\, P_X}
\ .
\label{polarcosdef}
\end{eqnarray}
Typically, the polar angles are very small and it is better to calculate
them in a numerically stable form from
\begin{eqnarray}
    \sin\theta_1 & = &  \frac{2\, \sqrt{D_1}}{s\, \beta\, P_1}
\nonumber\\
    \sin\theta_2 & = &  \frac{2\, \sqrt{D_3}}{s\, \beta\, P_2}
\nonumber\\
    \sin\theta_X & = &  \frac{2\, \sqrt{D_5}}{s\, \beta\, P_X}
\ .
\end{eqnarray}
Equations (\ref{polarcosdef}) are then only used to resolve the ambiguity 
$\theta_i \leftrightarrow \pi - \theta_i$. The quantity
$D_3$ is defined in (\ref{Didef}--\ref{sijdef}); $D_1$ is obtained from 
$D_3$ by the interchange $t_1 \leftrightarrow t_2$ and
$s_1 \leftrightarrow s_2$. The same interchange relates $D_2$, needed below, 
with $D_4$, given in (\ref{Didef}--\ref{sijdef}). Furthermore, we have:
\begin{eqnarray}
         D_5   & = &  D_1 + D_3 + 2\, D_6
\nonumber\\
         D_6  & = &  \frac{s}{8}\, \left[ 
             -(s-4\, m^2)\, (W^2-t_1-t_2)
                +(s_1-t_2-m^2)\, (s_2-t_1-m^2)+t_1\, t_2 \right]
\nonumber\\
     & &~       - \frac{m^2}{4}\, (s_1-m^2)\, (s_2-m^2)
\ .
\end{eqnarray}

The polar angles $\phi_1$ ($\phi_2$) between the $\e^+$ ($\e^-$) plane 
and the hadronic plane and the polar angle $\phi$ between the two lepton 
planes in the $\e^+\e^-$ c.m.s.\ are again best calculated using the 
numerically more stable form for the sinus function
\begin{eqnarray}
    \cos\phi & = &  \frac{D_6}{\sqrt{D_1\, D_3}}
\nonumber\\
    \sin\phi & = &  \frac{s\, \beta\, \sqrt{-\Delta_4}}{2\, \sqrt{D_1\, D_3}}
\nonumber\\
    \sin\phi_1 & = &  \frac{2\, \sqrt{-\Delta_4}}
              {s\, \beta\, P_X\, \sin\theta_X\, P_1\, \sin\theta_1}
\nonumber\\
    \sin\phi_2 & = &  \frac{2\, \sqrt{-\Delta_4}}
              {s\, \beta\, P_X\, \sin\theta_X\, P_2\, \sin\theta_2}
\nonumber\\
    \cos\phi_1 & = &  \frac{D_1+D_6}{\sqrt{D_1\, D_5}}
\nonumber\\
    \cos\phi_2 & = &  \frac{D_3+D_6}{\sqrt{D_3\, D_5}}
 = \frac{ \sqrt{D_3} + \sqrt{D_1}\, \cos\phi}
   {\sqrt{D_3 + D_1 + 2\, \sqrt{D_1\, D_3}\, \cos\phi} }
\ .
\end{eqnarray}

An expression for the azimuthal angle between the lepton planes in the 
$\gamma\gamma$ c.m.s.\ can be deduced from the formulas given in 
\cite{Budnev}: 
\begin{equation}
  \cos\tilde{\phi} = \frac{-2\, s+u_1+u_2-\nu+4\, m^2 + \nu(u_2-\nu)
  \, (u_1-\nu)/(K^2\, W^2)}{\sqrt{ t_1 \, t_2\,
 \left( 2\, \rho_1^{++}-2\right)\, \left( 2\, \rho_2^{++}-2\right)} }
\ .
\label{phibudnev}
\end{equation}
Numerically more stable is the following form
\begin{equation}
\sin\tilde{\phi}  =  \frac{K\, W\, \sqrt{- \Delta_4}}
         {\sqrt{D_2\, D_4}}
\qquad {\mathrm{and}} \qquad
\cos\tilde{\phi}  =  \frac{D_7}{\sqrt{D_2\, D_4}}
\ ,
\label{phistable}
\end{equation}
where
\begin{eqnarray}
%16\,\sqrt{D_2\, D_4}\, \cos\tilde{\phi} & = & 
16\, D_7 & = & 
2\,W^{2}\left (s_1\,s_2-sW^{2}\right )
%\nonumber\\
%& &~
-\, 2\,t_1\,\left (-t_1\,s_1+st_1+s_1\,s_2+
W^{2}s_1-2\,sW^{2}\right )
\nonumber\\
& &~
-\, 2\,t_2\,\left (-t_2\,s_2+t_2\,s+s_1\,s_2
-2\,sW^{2}+s_2\,W^{2}\right )
%\nonumber\\
%& &~
+\, 2\,t_1\,t_2\,\left (-s_1+2\,s+2\,W^{2}-s_2
\right )
\nonumber\\
& &~
-\, 2\,{m}^{2}\left ({m}^{2}t_2-{t_2}^{2}-{m}^{2}W^{2}
+{m}^{2}t_1-2\,W^{4}-{t_1}^{2}+W^{2}s_1+2\,t_1\,t_2
\right. 
\nonumber\\
& &~ \left. 
+\, 3\,t_1\,W^{2}-t_1\,s_1+3\,t_2\,W^{2}-
t_2\,s_2-t_2\,s_1+s_2\,W^{2}-t_1\,s_2
\right )
\ .
\end{eqnarray}

A numerically stable relation between $\phi$ and $\tilde{\phi}$ 
at $-t_i$, $m^2 \ll W^2$ is provided by 
\begin{eqnarray}
4\, s^2 \, D_2 & = & 4\, s_2^2\, D_3
%\nonumber\\ & &~ 
+\, 2\,t_2\,{s}^{2}r_2\,\cos\phi\,s_2
%\nonumber\\ & &~
-\, 2\,t_1\,t_2\,ss_2\,\left (s-s_1\right )
\nonumber\\ & &~
-\, st_1\,t_2\,\left (-2\,t_1\,t_2-st_1+t_1
\,s_1+3\,t_2\,s_2-t_2\,s+2\,r_2\,\cos\phi
\,s\right )
\nonumber\\ & &~
-\, 4\,{m}^{2}sr_2\,\cos\phi\,s_1\,s_2
%\nonumber\\ & &~
+\, O\left( m^4\, s_1^2\, s_2^2/s, m^2\, t_i\, s\, s_1\, s_2 \right)
%-{\frac {\left (4\,{s_2}^{2}{s_1}^{2}\sin^2\phi-8\,
%{s_2}^{2}{s_1}^{2}\right ){m}^{6}}{{s}^{2}}}
%\nonumber\\ & &~
%-{\frac {\left (-
%4\,{s}^{2}t_1\,{s_1}^{2}+4\,{s}^{2}{s_2}^{2}t_2\,
%\sin^2\phi-4\,{s}^{2}{s_2}^{2}t_2-4\,s{t_1}^{2}
%{s_1}^{2}+4\,s{s_2}^{2}{t_2}^{2}\sin^2\phi-4\,s
%{s_2}^{2}t_2\,s_1\,\sin^2\phi+4\,s_2\,{s}^{
%2}t_2\,s_1+4\,s{t_1}^{2}{s_1}^{2}{{\it sinphi}}^{2
%}-8\,r_2\,\cos\phi\,{s}^{2}s_2\,s_1-4\,ss_2\,
%t_1\,{s_1}^{2}\sin^2\phi+8\,ss_2\,t_1\,
%t_2\,s_1+4\,{s}^{2}t_1\,{s_1}^{2}\sin^2\phi
%-4\,s{s_2}^{2}{t_2}^{2}+4\,s{s_2}^{2}{s_1}^{2}
%\right ){m}^{4}}{{s}^{2}}}
%\nonumber\\ & &~
%+\left (-{\frac {-{t_2}^{2}{s}^{4}-4\,
%t_1\,t_2\,{s}^{4}+2\,s_2\,{t_2}^{2}{s}^{3}-{s}^{2}
%{t_1}^{2}{s_1}^{2}-{s}^{2}{s_2}^{2}{t_2}^{2}-8\,
%t_1\,{s}^{3}{t_2}^{2}+6\,{s}^{3}t_1\,t_2\,s_1-
%2\,s_2\,{s}^{3}t_2\,s_1+4\,s_2\,t_1\,t_2
%\,{s}^{3}+2\,{s}^{2}s_2\,t_1\,{s_1}^{2}-8\,{s}^{2}
%s_2\,t_1\,t_2\,s_1+8\,{s}^{2}t_1\,s_2\,
%{t_2}^{2}+8\,{t_1}^{2}{s}^{2}t_2\,s_1-4\,{t_1}^{2}
%{s}^{3}t_2-8\,{t_1}^{2}{s}^{2}{t_2}^{2}+4\,{t_1}^{2
%}{s}^{3}t_2\,\sin^2\phi+4\,t_1\,{s}^{3}{t_2}^{2
%}\sin^2\phi-4\,{s}^{2}t_1\,s_2\,{t_2}^{2}
%\sin^2\phi-4\,{s}^{3}t_1\,t_2\,s_1\,{{\it sinphi}}^{2
%}+4\,{s}^{2}s_2\,t_1\,t_2\,s_1\,\sin^2\phi
%-4\,s_2\,t_1\,t_2\,{s}^{3}\sin^2\phi-4\,
%{t_1}^{2}{s}^{2}t_2\,s_1\,\sin^2\phi+4\,t_1\,
%t_2\,{s}^{4}\sin^2\phi+4\,{t_1}^{2}{s}^{2}{t_2}^
%{2}\sin^2\phi+4\,r_2\,\cos\phi\,{s}^{3}s_1\,
%s_2+8\,r_2\,\cos\phi\,{s}^{3}t_1\,t_2-4\,
%r_2\,\cos\phi\,{s}^{3}t_1\,s_1-4\,r_2\,\cos\phi
%\,{s}^{3}s_2\,t_2+4\,r_2\,\cos\phi\,{s}^{4}
%t_2}{{s}^{2}}}+4\,sr_2\,\cos\phi\,s_1\,s_2\right )
%{m}^{2}
\ ,
\label{numeristableone}
\end{eqnarray}
an analogous expression for $D_4$, and
\begin{eqnarray}
%s N(\cos\tilde{\phi}) & = & 16\, W^2\, D_6 
%4s\sqrt{G_2 G_2'}\cos\tilde{\phi} & = & 4 W^2 \sqrt{G1 G_1'}\cos\phi
16\, s\, \sqrt{D_2 D_4}\cos\tilde{\phi} & = & 16\, W^2 \sqrt{D_1 D_3}\cos\phi
%\nonumber\\ & &~ 
-\, 4\,t_2\,{s}^{2}r_2\,\cos\phi
%\nonumber\\ & &~
-\, 4\,t_1\,{s}^{2}r_2\,\cos\phi
\nonumber\\ & &~
+\, 4\,t_1\,t_2\,s\left (-s_1+2\,s-s_2+t_1
+t_2\right )
%\nonumber\\ & &~
+\, 8\,{m}^{2}\cos\phi\,r_2\,s_1\,s_2
\nonumber\\ & &~ +
\frac {2\,{m}^{2}t_2}{s}\,\left (-t_2\,{s}^{2}+4\,st_2\,
s_2-2\,t_2\,{s_2}^{2}-4\,ss_1\,s_2+2\,s_1
\,{s_2}^{2}
\right. \nonumber\\ & &~~ \left. 
-\, 8\,r_2\,\cos\phi\,ss_2
+  {s}^{2}s_1+
{s}^{2}s_2+8\,r_2\,\cos\phi\,{s}^{2}\right )
\nonumber\\ & &~ 
+\, \frac {2\,{m}^{2}t_1}{s}\,\left (-t_1\,{s}^{2}+4\,t_1\,s
s_1-2\,t_1\,{s_1}^{2}-4\,ss_1\,s_2+2\,
{s_1}^{2}s_2
\right. \nonumber\\ & &~~ \left. 
+\, 8\,r_2\,\cos\phi\,{s}^{2}
- 8\,r_2\,
\cos\phi\,ss_1+{s}^{2}s_1+{s}^{2}s_2\right )
\nonumber\\ & &~
+\, O\left( m^4\, s_i^2\, s_j/s, m^2\, t_1\, t_2\, s\right)
\ ,
\label{numeristabletwo}
\end{eqnarray}
where
\begin{equation}
r_2^2 = \left[ m^2\, \left({s_1\over s}\right)^2 + 
       t_2\, \left(1+{t_2\over s}-{s_1\over s}\right) \right] \,
        \left[ m^2\, \left({s_2\over s}\right)^2 + 
       t_1\, \left(1+{t_1\over s}-{s_2\over s}\right) \right] 
\ .
\end{equation}
For $m\rightarrow 0$ and $t_i/W^2 \rightarrow 0$, 
(\ref{numeristableone}) and (\ref{numeristabletwo}) lead to the
approximate relation (\ref{phitildeapprox}). 

The four-momenta are now given by
\begin{eqnarray}
  p_a & =  & \frac{1}{2}\, \sqrt{s}\, 
   (1, 0, 0, -\beta)
\nonumber\\
  p_b & =  & \frac{1}{2}\, \sqrt{s}\, 
   (1, 0, 0, \beta)
\nonumber\\
  p_1 & =  & (E_1, - P_1\, \sin\theta_1\, \cos\phi_1,
  - P_1\, \sin\theta_1\, \sin\phi_1, - P_1\, \cos\theta_1)
\nonumber\\
  p_2 & =  & (E_2, - P_2\, \sin\theta_2\, \cos\phi_2,
  P_2\, \sin\theta_2\, \sin\phi_2, P_2\, \cos\theta_2)
\nonumber\\
  p_X & = & (E_X, P_X\, \sin\theta_X, 0, P_X\, \cos\theta_X)
\ .
\end{eqnarray}

%----------------------------------------------------------------

\section{Experimental cuts}
\label{sec:cuts}
\begin{figure}
\epsfig{file=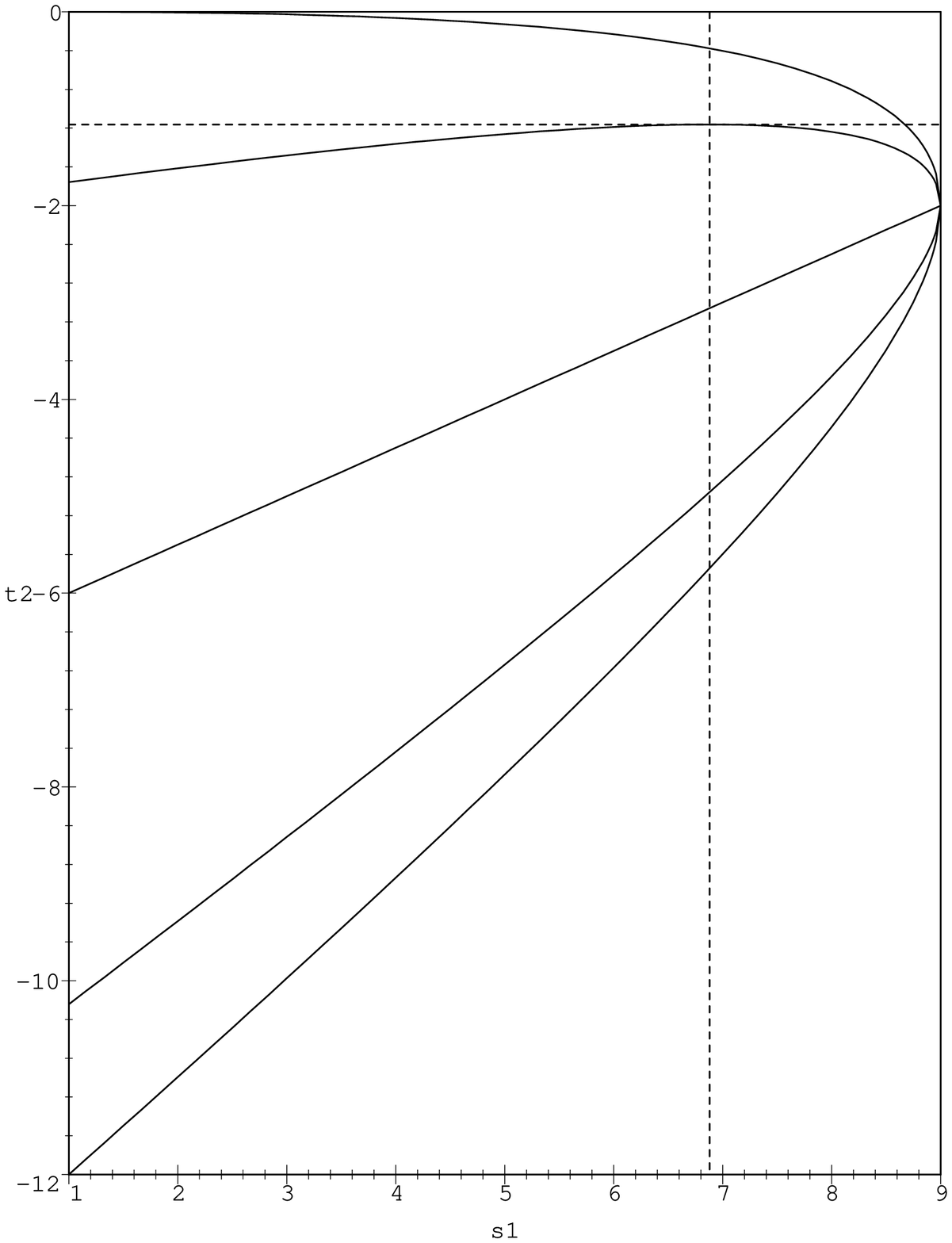,width=\textwidth}
\caption[]{Phase space in the variables $(t_2,s_1)$ for $\sqrt{s}=4$
and $m = 1$. The solid lines correspond to
$\theta_2 = 0$, $\pi/4$, $\pi/2$, $3\pi/4$, and $\pi$ (from $t_2=0$
to $t_2=-12$ at $s_1=1$). The dashed lines are $s_1 = \hat{s}_1$
and $t_2 = \hat{t}_2$ at $\theta_2=\pi/4$.
\label{fig:phasespace}}
\end{figure}

If cuts on the angle $\theta_2$ and the energy $E_2$ of the 
scattered electron are applied, the $(s_1,t_2)$-integration 
region shrinks as follows (see Fig.~\ref{fig:phasespace}): 
\begin{eqnarray}
  s_{1\low} & = & \min\, \left\{ (m+W)^2, m^2 + s\,
  \left( 1 - \frac{2\, E_{2\max}}{\sqrt{s}} \right) \right\}
\nonumber\\
  s_{1\upp} & = & \max\, \left\{ (\sqrt{s}-m)^2, m^2 + s\,
  \left( 1 - \frac{2\, E_{2\min}}{\sqrt{s}}\right) \right\}
\end{eqnarray}
and
\begin{equation}
  T_2(s_1,\theta_{2\max}) < t_2 < T_2(s_1,\theta_{2\min})
\ ,
\end{equation}
where
\begin{eqnarray}
  T_2(s_1,\theta_2) & = & \frac{1}{2}\, \left(
         3\, m^2 - s + s_1 + \beta\, \cos\theta_2\, 
              \sqrt{\lambda(s,s_1,m^2)} \right)
\nonumber\\
%  & = & \frac{m^2\, (s_1 - m^2)^2/s + \beta^2\, \sin^2\theta_2\,
%    \lambda(s,s_1,m^2)/4}{\left(
%         3\, m^2 - s + s_1 - \beta\, \cos\theta_2\, 
%               \sqrt{\lambda(s,s_1,m^2)}   \right)/2}
  & = & 
 - \frac{2\, m^2\, (s_1 - m^2)^2}{s\, \left[ \beta\, \lambda^{1/2}
  (s,s_1,m^2) + s - s_1 - 3\, m^2 \right]} - \beta\, 
  \lambda^{1/2}(s,s_1,m^2)\, \sin^2\frac{\theta_2}{2}
\nonumber\\
  & \rightarrow & - \left\{ \frac{m^2\, x_2^2}{1-x_2} + s\, (1-x_2)
  \, \sin^2\frac{\theta_2}{2}  \right\}
\ .
\end{eqnarray}
The approximate form holds for $m^2 \ll s_1$ and a small 
angle $\theta_2$ and is used in (\ref{tiapproxlimits}).

If, as in our case, $t_2$ is the outer integration, then its lower limit
becomes
\begin{equation}
  t_{2\min} = \min\, \left\{ T_2(s_{1\upp},\theta_{2\max}),  
            T_2(s_{1\low},\theta_{2\max})  \right\}
\ ,
\label{t2mincut}
\end{equation}  
while the upper limit is more complicated
\begin{eqnarray}
  t_{2\max}  = & T_2(s_{1\upp},\theta_{2\min}) 
           \qquad &  \hat{s}_1 > s_{1\upp}
\nonumber\\
        = & T_2(s_{1\low},\theta_{2\min}) 
           \qquad &  \hat{s}_1 < s_{1\low}
\nonumber\\
        = & \hat{t}_2 \qquad & s_{1\low} < \hat{s}_1 < s_{1\upp} 
\ ,
\label{t2maxcut}
\end{eqnarray}
where
\begin{eqnarray}
  \hat{s}_1 & = & s + m^2 - \frac{2\,m\, \sqrt{s}}{\sqrt{X}}
\nonumber\\
         & = & m^2 + \frac{s\, \beta^2\, \sin^2\theta_2}
          {X\, \left( 1 + 2\, m\, /\sqrt{s\, X} \right) }
\nonumber\\
  \hat{t}_2 & = & 2\, m^2 - m\, \sqrt{s\, X} \qquad (\theta_2 < \pi/2) 
\nonumber\\
            & = & 2\, m^2 - m\, \sqrt{s}\, 
           \frac{1 + \beta^2\, \cos^2\theta_2}
                {\sqrt{X}}  \qquad (\theta_2 > \pi/2) 
\nonumber\\
  X & = & \frac{4\, m^2}{s} + \beta^2\, \sin^2\theta_2
\ .
\label{cuth1}    
\end{eqnarray}
The $s_1$-integration range is a rather complicated function of
$t_2$ and may even consist of two separated ranges  
(Fig.~\ref{fig:phasespace}). Moreover, the $s_1$-integration 
range is affected by $t_1$ and cuts on $E_1$ and $\theta_1$.
Then it is better to use the Monte Carlo method. In any case, 
since the $t_i$ integration are the most singular ones, the most 
important constraints are taken into account through (\ref{t2mincut})
and (\ref{t2maxcut}) and the analogous formulas for $t_1$.
%%%%%%%%%%%%%%%%%%%%%%%%%%%%%%%%%%%%%%%%%%%%%%%%%%%%%%%%%%%%%%%%%%%%%%%%%%%

\renewcommand{\arraystretch}{1.0}

%\clearpage
\section{Details of the program}
\label{sec:program}
\subsection{Common blocks}
The user can decide whether to keep $t_2$ at a fixed, user-defined value 
or to integrate over $t_2$, i.e.\ to calculate (\ref{crossee}) 
or (\ref{DIScross}).
In the case of integration over all variables, the user can choose between 
the exact or an approximate treatment (\ref{approxcross}) of the kinematics.
\begin{verbatim}
      Common /ggLapp/ t2user,iapprx,ivegas,iwaght
\end{verbatim}
\begin{tabular}{ll}
 \verb+ t2user + & Fixed value of $t_2$ chosen by user for 
     \verb+ iapprx = 1. +
\\
 \verb+ iapprx + & $=1$: $t_2$ is kept fixed at the user value;
\\
                 & $=2$: approximate kinematics is used, 
                 $t_2$ is integrated over;
\\
                 & $=0$: all variables are integrated using exact kinematics.
\\
 \verb+ ivegas + & $=1$: VEGAS integration;
\\
                 & $=0$: Simple integration.
\\
 \verb+ ivegas + & $=1$: Unweighted events, i.e.\ \verb+ Weight+ $=1$;
\\               & $=0$: Weighted events
\\[2ex]
\end{tabular}

\noindent
Cuts on the scattered leptons are set in
\begin{verbatim}
      Common /ggLcut/th1min,th1max,E1min,E1max,th2min,th2max,E2min,E2max
\end{verbatim}
\begin{tabular}{ll}
 \verb+ th1min,th1max + & Minimum and maximum scattering angles of  
  scattered $\e^+$ 
\\ & w.r.t.\ direction of incident $\e^+$.
\\    
 \verb+ th2min,th2max + & Minimum and maximum scattering angles of  
  scattered $\e^-$ 
\\ & w.r.t.\ direction of incident $\e^-$.
\\ & Tighter cuts should be applied to the $\e^-$.
\\    
 \verb+ E1min,E1max + & Minimum and maximum energies of
  scattered $\e^+$.
\\    
 \verb+ E2min,E2max + & Minimum and maximum energies of
  scattered $\e^-$.
\\[2ex]
\end{tabular}

\noindent
Models for the $\gamma^\star\gamma^\star$ cross sections and their
parameters are chosen in
\begin{verbatim}
      Common /ggLmod/ imodel
\end{verbatim}
\begin{tabular}{ll}
%\verb+ imodel + $=1$, $2$, $0$ & Models according to 
%(\ref{GVMDform}), (\ref{VMDcform}), (\ref{rhoform});
%\\
\verb+ imodel+ $=~1\qquad$ & GVMD model (\ref{GVMDform})
\\
\verb+ imodel+ $=~2$ & VMDc model (\ref{VMDcform})
\\
\verb+ imodel+ $=20$ & $\rho$-pole model (\ref{rhoform})
\\
\verb+ imodel+ $=~0$ & $\rho$-pole model (\ref{rhoform}) with $h_S(Q^2)=0$
\\
\verb+ imodel+ $=~3$ & Exact cross section for lepton-pair production.
\end{tabular}

\begin{verbatim}
      Common /ggLhad/ r,xi,m1s,m2s,rrho,romeg,rphi,rc,mrhos,
     &     momegs,mphis,mzeros
\end{verbatim}
Parameters for (\ref{GVMDform}--\ref{rhoform}): $r$, $\xi$, $m_1^2$, 
$m_2^2$, $r_\rho$, $r_\omega$, $r_\phi$, $r_c$, $m_\rho^2$, 
$m_\omega^2$, $m_\phi^2$, $m_0^2$.\hfill\\[1ex]

\noindent
The integration variables and the particle momenta are stored in
\begin{verbatim}
      Common /ggLvar/ 
     &yar(4),t2,t1,s1,s2,E1,E2,EX,P1,P2,PX,th1,th2,thX,phi1,phi2,phi,pht
\end{verbatim}
\begin{tabular}{ll}
  \verb+ yar(i) + & Integration variables for VEGAS.
\\
  \verb+ t2,t1,s1,s2 + & Invariants $t_2$, $t_1$, $s_1$, $s_2$.
\\
  \verb+ E1,E2,EX + & Energies $E_1$, $E_2$, $E_X$.
\\
  \verb+ P1,P2,PX + & Three-momenta $P_1$, $P_2$, $P_X$.
\\
  \verb+ th1,th2,thX + & Polar angles $\theta_1$, $\theta_2$, $\theta_X$.
\\
  \verb+ phi1,phi2,phi,pht + & Azimuthal angles 
      $\phi_1$, $\phi_2$, $\phi$, $\tilde{\phi}$.
\end{tabular}

\begin{verbatim}
      Common /ggLvec/ mntum(7,5)
\end{verbatim}
Particle four-momenta \verb+ mntum(i,k): + $k=1 \ldots 5$ for
$p_x$, $p_y$, $p_z$, $E$, ${\mrm{sign}}(p^2) \times \sqrt{|p^2|}$;
$i=1 \ldots 7$ for incident $\e^+$, incident $\e^-$, 
photon from $\e^+$, photon from $\e^-$, 
scattered $\e^+$, scattered $\e^-$, hadronic system $X$.
\hfill\\[1ex]      

\noindent
Parameter for the simple integration and 
results of the integration and event generation are stored in
\begin{verbatim}
      Common /ggLuno/ cross,error,Fmax,Fmin,Weight,npts,nzero,ntrial
\end{verbatim}
\begin{tabular}{ll}
 \verb+ cross + & Estimate of luminosity.
\\
 \verb+ error + & Estimate of error on luminosity.
\\
   \verb+ Fmax + &  Maximum function value, calculated in 
       \verb+ggLcrs+; 
\\ & checked in \verb+ggLgen+.
\\
   \verb+ Fmin + &  Minimum function value, calculated in 
       \verb+ ggLuF. + 
\\
 \verb+ Weight + & Weight if weighted events requested.
\\
   \verb+ npts + &  Number of function evaluations for simple integration.
\\
 \verb+ nzero + & Number of cases where function was put to zero 
  in \verb+ ggLuF+ \\
  & because it failed the cuts; \\
  & initialized to zero in \verb+ ggLcrs+, \verb+ ggLgen+.
\\  \verb+ ntrail + & Number of trials necessary in \verb+ ggLgen + 
  to generate an event;
\\
  & incremented by each call.
\\[2ex]
\end{tabular}

\noindent
Parameters for the VEGAS integration are set in
\begin{verbatim}
      Common /ggLvg1/ xl(10),xu(10),acc,ndim,nfcall,itmx,nprn
\end{verbatim}
\begin{tabular}{ll}
 \verb+ acc + & VEGAS accuracy (Default (D): $10^{-4}$).
\\      
 \verb+ ndim + & Number of integration variables (D: $4$).
\\        
 \verb+ nfcall + & Maximum number of function calls 
       per iteration for VEGAS (D: $10^5$).
\\        
 \verb+ itmx + & Number of iterations for VEGAS (D: $4$).
\\         
 \verb+ nprn + & Print flag for VEGAS (D: $2$).
\\[2ex]
\end{tabular}

\noindent
Additional common blocks
\begin{verbatim}
      Common /ggLprm/ s,roots,Whad,m,Pi,alem
\end{verbatim}
\begin{tabular}{ll}
 \verb+ s + & Overall c.m.\ energy square $s$.
\\
 \verb+ roots + & Overall c.m.\ energy $\sqrt{s}$ (twice the beam energy), 
\\ & set by user through call to \verb+ ggLcrs. +
\\
 \verb+ Whad + & Hadronic mass $W$, set by user through call 
  to \verb+ ggLcrs. +
\\
 \verb+ m + & Electron mass (D: $511\,$keV).
\\
 \verb+ Pi + & $\pi$
\\
 \verb+ alem + & $\alpha_{{\mrm{em}}}$ (D: $1/137$).
\end{tabular}

\begin{verbatim}
      Common/ggLvg2/XI(50,10),SI,SI2,SWGT,SCHI,NDO,IT

      Common /ggLerr/ 
     &  it1,iD1,iD3,iD5,itX,iph,ip1,ip2,ia1,ia2,ia3,ia4,ie1,ie2,ipt,is

      Block Data ggLblk
\end{verbatim}
%      Common /PAWC/ HMEMOR(100000)
%\end{verbatim}

\subsection{Subroutines}
\begin{tabular}{ll}
\verb+ ggLcrs(rs,W) + & Integrates $\d\sigma/\d \tau$ and 
  finds \verb+ Fmax+; 
%   or $\d\sigma/\d\tau\d t_2$ within cuts defined in\\ & \verb+ ggLcut +. 
  \verb+ rs+ $=\sqrt{s}$, \verb+ W+ $=W$
\\[1ex]
\verb+ ggLmom + & Builds up four-momenta.
\\[1ex]
\verb+ ggLprt + & Prints four-momenta and checks momentum sum.
%\\[1ex]
%\verb+ User(n) + & User routine:
%\\ &
% performs user action after generation of an event.
%\\[1ex]
%\verb+ ggLuin + & Finds the maximum function value.
\\[1ex]
\verb+ ggLgen(Flag) + & Generates one event;
%\\ & \verb+cwgt=T+ to request unweighted events;
\\ &
\verb+Flag=F+ if a new maximum is found; then it is advisable 
\\ &
to restart event generation with adjusted maximum.
% distributions could be wrong.
\end{tabular}

\subsection{Double-precision functions}
\begin{tabular}{ll}
\verb+ ggLint(W2,m2,Q1s,Q2s,s1,s2,phi,s) + & $\Sigma$ as defined in 
(\ref{Sigmadef}).
\\[1ex]
\end{tabular}

\noindent
\begin{tabular}{ll}
\verb+ ggLuF(xar,wgt) + & $F(x_i)$ as defined in (\ref{Fdef}).
\\[1ex]
\verb+ ggLhTT(W2,Q1s,Q2s) + & $\sigma_{TT}(W^2,Q_1^2,Q_2^2)$
%calculated as in (\ref{sigapprox}).
\\[1ex]
\verb+ ggLhTS(W2,Q1s,Q2s) + & $\sigma_{TS}(W^2,Q_1^2,Q_2^2)$
%calculated as in (\ref{sigapprox}).
\\[1ex]
\verb+ ggLhSS(W2,Q1s,Q2s) + & $\sigma_{SS}(W^2,Q_1^2,Q_2^2)$
%calculated as in (\ref{sigapprox}).
\\[1ex]
\verb+ ggLrTS(W2,Q1s,Q2s) + & $\tau_{TS}(W^2,Q_1^2,Q_2^2)$
%put to zero.
\\[1ex]
\verb+ ggLrTT(W2,Q1s,Q2s) + & $\tau_{TT}(W^2,Q_1^2,Q_2^2)$
%put to zero.
\\[1ex]
\verb+ ggLhT(Qs) + & $h_{T}(Q^2)$
%as given by the models (\ref{GVMDform}--\ref{rhoform}).
\\[1ex]
\verb+ ggLhS(Qs) + & $h_{S}(Q^2)$
%as given by the models (\ref{GVMDform}--\ref{rhoform}).
\\[1ex]
\verb+ ggLgg(W2) + & $\sigma_{\gamma\gamma}(W^2)$
%put to one.
\\[1ex]
\verb+ ggLuG(z) + & Makes the variable transformation from $x_i$ in 
(\ref{Fdef}) to 
\\ & those used by the simple or VEGAS integration.
\end{tabular}

\clearpage
\subsection{Excerpt from the demonstration program}
\begin{verbatim}
* Initialize the random number generator RanLux
      Call rLuxGo(3,314159265,0,0)
*
* Initialize GALUGA; get luminosity within cuts
      Call ggLcrs(rs,W)
*
* Initialize plotting
      Call User(0)
*
* Timing:
      Call Timex(time1)
      Call rLuxGo(3,314159265,0,0)
*
* Event loop
      Do 10 i=1,Nev
         Call ggLgen(Flag)
         If(.not.Flag) Write(6,*) 'Caution: new maximum'
*
* Calculate 4-momenta
         Call ggLmom
*
* Display first 3 events
         If(i.le.3) call ggLprt
*
* Fill histrograms
         Call User(1)
 10   Continue
*
      Call Timex(Time2)
      Write(6,300) Nev,Time2-Time1,(Time2-Time1)/real(Nev),
     &     iwaght,ntrial,nzero,Fmax
*
* Finalize plotting
      Call User(-Nev)
*
 300  Format(/,3x,'time to generate ',I8,' events is       ',E12.5,/,
     &3x,'resulting in an average time per event of ',E12.5,/,
     &3x,'unweighted events requested if 1:         ',I8,/,
     &3x,'the number of trials was:                 ',I8,/,
     &3x,'the number of zero f was:                 ',I8,/,
     &3x,'the (new) maximum f value was:            ',E12.5)
*
      Stop
\end{verbatim}
 
%%%%%%%%%%%%%%%%%%%%%%%%%%%%%%%%%%%%%%%%%%%%%%%%%%%%%%%%%%%%%%%%%%%%%%%%%%%

\clearpage
%---------------------------------------------------
%----------------------------------------------------------------
%---------------------------------------------------
%---------------------------------------------------
%---------------------------------------------------
%---------------------------------------------------
%---------------------------------------------------
%---------------------------------------------------
%---------------------------------------------------
%----------------------------------------------------------------
%%%%%%%%%%%%%%%%%%%%%%%%%%%%%%%%%%%%%%%%%%%%%%%%%%%%%%%%%%%%%%%%

\clearpage
\begin{thebibliography}{99}
%
\bibitem{VEGAS}
G.P.\ Lepage, {\it J.\ Comp.\ Phys.\ } {\bf 27} (1978) 192
\bibitem{RANLUX}
F.\ James, ``RANLUX: A Fortran implementation of the high-quality 
pseudorandom number generator of L\"{u}scher'', 
CERN Program Library V115, \CPC{79} (1994) 111;\hfill\\
M.\ L\"{u}scher, \CPC{79} (1994), 100
\bibitem{HBOOK}
R.\ Brun and D.\ Lienart, ``HBOOK User Guide -- Version~4'', 
CERN Program Library Y250, 1988
\bibitem{DATIME}
J. Harms et al., ``DATIME: Job Time and Date'', 
CERN Program Library Z007, 1991
\bibitem{LEP2}
Report on `$\gamma\gamma$ Physics', conveners P.\ Aurenche and G.A.\ 
Schuler, in Proc.\ Physics at LEP2, eds.\ G.\ Altarelli, T.\ Sj\"ostrand
and F.\ Zwirner (CERN 96-01, Geneva, 1996), Vol.~1, p.~291; \hep{9601317}
\bibitem{Budnev}
V.M.\ Budnev et al., \PRC{15} (1975) 181, and references therein
\bibitem{vermaseren}
J.A.M.\ Vermaseren, \NPB{229} (1983) 347
\bibitem{diag36}
``DIAG36'',
F.A.\ Berends, P.H.\ Daverveldt and R.\ Kleiss, \NPB{253} (1985) 421; 
\CPC{40} (1986) 271, 285, and 309
\bibitem{Diberder}
``FERMISV'',
F.\ Le Diberder, J.\ Hilgert and R.\ Kleiss, \CPC{75} (1993) 191
\bibitem{herwig}
``HERWIG'', G.\ Marchesini et al., \CPC{67} (1992) 465
\bibitem{pythia}
``PYTHIA'', T.\ Sj\"ostrand, \CPC{82} (1994) 74; 
Lund University report LU-TP-95-20 (1995)
\bibitem{phojet}
``PHOJET'', 
R.\ Engel and J.\ Ranft, \PRD{54} (1996) 4244;\hfill\\
R.\ Engel, \ZPC{66} (1995) 203
\bibitem{minijet}
``MINIJET'', A.\ Miyamoto and H.\ Hayashii, \CPC{96} (1996) 87
\bibitem{ggps1}
``GGPS1/2'', T.\ Munehisa, K.\ Kato and D.\ Perret--Gallix, 
in same Proc.\ as in ref.\ \cite{LEP2}, 
Vol.~2, p.~211;
\hfill\\
T.\ Munehisa, P.\ Aurenche, M.\ Fontannaz and Y.\ Shimizu, 
preprint KEK CP 032 (1995), \hep{9507339}
%``QEDPS'',
%T.\ Munehisa, J.\ Fujimoto, Y.\ Kurihara and Y.\ Shimizu,  
%preprint KEK-CP-034 (1995), \hep{9603322}
%\bibitem{grc4f}
%``GRC4F'', J.\ Fujimoto et al., KEK-CP-046 (1996)
%\hep{9605312}
\bibitem{gghv01}
``GGHV01'', M.\ Kr\"amer, P.\ Zerwas, J.\ Zunft and A.\ Finch, 
in same Proc.\ as in ref.\ \cite{LEP2}, 
Vol.~2, p.~210
\bibitem{GAS}
G.A.\ Schuler, preprint CERN-TH/96-297, \hep{9610406}
\bibitem{twogam}
``TWOGAM'', S.\ Nova, A.\ Olshevski and T.\ Todorov, DELPHI Note 90-35
(1990)
\bibitem{twogen}
``TWOGEN'', A.\ Buijs, W.G.J.\ Langeveld, M.H.\ Lehto and D.J.\ Miller,
\CPC{79} (1994) 523
\bibitem{SaS}
G.A.\ Schuler and T.\ Sj\"ostrand, preprint CERN-TH-96-119, 
\hep{9605240}; \NPB{407} (1993) 539
\bibitem{Byckling}
E.\ Byckling and K.\ Kajantie, ``Particle kinematics'', 
(John Wiley \& Sons, New York, 1973);
\hfill\\
K.\ Kajantie and P.\ Lindblom, {\it Phys.\ Rev.\ } {\bf 175} (1968) 2203
\bibitem{gagaMC}
Report on `$\gamma\gamma$ Physics', conveners L.\ L\"onnblad and M.\ 
Seymour, 
in same Proc.\ as in ref.\ \cite{LEP2}, 
Vol.~2, p.~187; \hep{9512371}
\bibitem{Arteaga}
N.\ Arteaga, C.\ Carimalo, P.\ Kessler, S.\ Ong and O.\ Panella, 
\PRD{52} (1995) 4920
\bibitem{Bezrukov}
 L.B.\ Bezrukov and E.V.\ Bugaev, {\it Sov.\ J.\ Nucl.\ Phys.\ }
 {\bf 32} (1980) 847
\bibitem{Sakurai}
J.J.\ Sakurai and D.\ Schildknecht, \PLB{40} (1972) 121
%\bibitem{PDG}
%Particle Data Group, L.\ Montanet et al., \PRD{50} (1994) 1173
%\bibitem{Field}
%J.\ Field, \NPB{168} (1980) 477
%\bibitem{Bonneau}
%G.\ Bonneau, M.\ Gourdin and F.\ Martin, \NPB{54} (1973) 573
%\bibitem{HERA}
%H1 collab., S.\ Aid et al., \ZPC{69} (1995) 27;\hfill\\
%ZEUS collab., M.\ Derrick et al., \ZPC{63} (1994) 391;\hfill\\
%A.\ Levy, DESY-95-204, to appear in Proc.\ EPS HEP 95, Brussels, Belgium,
%1995, and references therein;\hfill\\
%Proc.\ Workshop on HERA Physics, Durham, 1995, J.\ Phys.\ {\bf G22}
%(1996);\hfill\\
%Proc.\ Photon '95, Sheffield, 1995, eds.\ D.J.\ Miller, S.L.\ 
%Cartwright and V.\ Khoze (World Scientific, Singapore, 1995)
%\bibitem{lowQ}
%H1 collab., paper pa02-070 submitted to the 28th Int.\ Conf.\
%on High Energy Physics, Warsaw, 1996; \hfill\\
%ZEUS collab., ibid., paper pa02-025
%\bibitem{gagastar}
%J.\ Bartels, A.\ DeRoeck and H.\ Lotter, DESY-96-168 , \hep{9608401}; 
%\hfill\\
%S.J.\ Brodsky, F.\ Hautmann and D.E.\ Soper,  SLAC-PUB-7218-T-E,
%\hep{9610260}
%\bibitem{Ridolfi}
%S.\ Frixione, M.L.\ Mangano, P.\ Nason and G.\ Ridolfi, 
% \PLB{319} (1993) 339
\end{thebibliography}
\end{document}